\documentclass{svjour2}                    
\smartqed  
\usepackage{graphicx}
\begin{document}

\title{Self-interacting quantum electron 
}


\author{Peter Leifer 
}


\institute{Peter Leifer \at
              Haatid, College for Science and Technology, 18123 Afula, Israel  \\
              Tel.: +972-4-6405580\\
              Fax: +972-4-6405576\\
              \email{leifer@bezeqint.net }           
           \and
}

\date{Received: date / Accepted: date}

\maketitle

\begin{abstract}
A model of a self-interacting quantum nonlocal Dirac electron has been proposed. Its dynamics was revealed by the projective representation of operators corresponding to spin/charge degrees of freedom. The energy-momentum field is described by the system of quasi-linear ``field-shell" PDE's following from the conservation law expressed by the affine parallel transport of the energy-momentum vector field in $CP(3)$. I discuss here traveling wave solutions of these equations and the ``off-shell" dispersion law asymptotically coinciding with the ``on-shell" de Broglie dispersion law.

\end{abstract}

\section{Introduction}
\label{intro}
Statistical analysis of the energy distribution is the basis of the black body
radiation \cite{Planck1} and the Einstein's hypothesis of the light emission and absorption \cite{Einstein_Q}. Success of Einstein hypothesis of photons, de Broglie wave concept of particles \cite{dB} and the Schr\"odinger equation for hydrogen atom \cite{Sch1} established so-called the corpuscular-wave duality of matter.
This conceptual line was logically finished by Dirac in his
method of the second quantization \cite{Dirac1}. This approach perfectly fits to
many-body weakly interacting quantum systems and it was assumed that the
``corpuscule-wave duality" is universal. This duality may be broken in strong interacting quantum systems and even for a single particle. Physically it is clear why: the quantum particle is self-interacting system and this interaction is at least of the order of its rest mass. Since the nature of the mass is an open problem we do not know the energy distribution in quantum particles up to now. Here I try to show a possible approach to this problem in the framework of simple model of self-interacting quantum electron with possible ``unparticle" excitations. The ``unparticle" sector of quantum excitations is intensively discussed now in the framework of effective QFT \cite{Georgi1,Georgi2,Gaete}.

I should note that Blochintzev about 60 years ago discussed the unparticle
sector in the framework of universality of wave - particle ``duality" for interacting quantum fields  \cite{Bl1,Bl2}. For such fields universality is generally broken. Namely the attempt to represent two interacting boson fields as the set of free quantum
oscillators leads to two types of oscillators: quantized and non-quantized. The second gives rise to the simple relation $g > \frac{m_1 m_2c^2}{h^2}$ between coupling the constant $g$  and masses $m_1$ and $m_2$ of two scalar fields. For such intensity of coupling we obtain a field with excitation states in two sectors: particle and
``unparticle". Furthermore, excitations in ``unparticle" sector have imaginary mass and they propagate with group velocity larger than $c$. For a self-interacting
scalar field of mass $m$ the intensity of self-interaction $g$ leads to breakdown of the universality of the wave - particle ``duality" if it is larger than the inverse square of the Compton wavelength: $g > \frac{m^2c^2}{h^2}=\frac{1}{\lambda^2_C}$.

Blochintzev's examples were oversimplified for clarity.
I would like to discuss here the self-interacting electron in the spirit of reaction
$e^-\to \mathcal{U} \to e^-$. In other words I propose to study the particle/unparticle sectors of matter in a wide range of momenta in order to solve the localization problem of the foundations of quantum physics.
In order to formulate a robust theory of self-interacting quantum ``particles", say, electron, one should analyse the quantum invariants and their relations to space-time symmetries.

The fundamental observation of quantum interference shows that variation in the quantum setup leads generally to deformation of interference patterns. Quantum formalism generally shows that two setups $S_1$ and $S_2$ generate two different amplitudes $|S_1>$ and $|S_2>$ of outcome event. There are an infinite number of different setups $S_1, S_2, ..., S_p,...$ and not only in the sense of different space-time position but  also in the sense of different parameters of fields, using devices, etc. Symmetries relative to the space-time transformations of whole setup have been studied in ordinary quantum theory. Such symmetries reflect, say, the
\emph{first order of relativity}:
the physics is same if any \emph{complete setup} is subject to kinematical shifts, rotations, boosts as a whole in single Minkowski space-time.

Further thinking leads to conclusion that there is a different type of symmetry (\emph{second order of relativity or ``super-relativity"} \cite{Le1,Le2,Le3}).
It may be formulated initially on the intuitive level as the invariance of physical properties of ``quantum particles" , i.e. their quantum numbers like mass, spin, charge, etc., underlying the two amplitudes $|S_1>, |S_2>$. Physical properties of electrons are the same in both setups $S_1$ and $S_2$ but they may be hidden in amplitudes of different outcomes. Presumably the invariant content of these properties may be kept if one makes the infinitesimal variation of some ``flexible quantum setup" that may be reached by small variation of some fields or adjustment of tuning devices.

Of course, all non-essential details of a real setup should be avoided in the problem where one seeks the invariant properties of quantum objects underlying the generating amplitudes. Otherwise we will trapped in the Bohr's tenet of ``classical language" leading to mixture of quantum and classical language that is the obstacle for building pure quantum model. This is why Fock's principle of ``relativity to measuring device" \cite{Fock} and ``functional relativity" \cite{Kryukov1,Kryukov2} could not be realized in full measure, since there is no and could not be a good mathematical quantum model for classical setup. Therefore a model of ``flexible pure quantum setup" with a possibility for infinitesimal variation of some parameter (in my model parameters of $SU(N)$ ) should be built.
In order to do it one needs to find invariant laws of quantum motions and to provide their classification. If we limit ourself by unitary finite dimension dynamics then the group $SU(N)$ acting in $C^N$ may be used. This approach was developed as a framework of \emph{``local functional relativity"} or \emph{``super-relativity"} \cite{Le1,Le2,Le3}. This is the actual physical reason why I use vector fields on
$CP(N-1)$ playing the role of local dynamical variables (LDV's) in order to build
\emph{flexible quantum reference frame} \cite{Le4}. All arguments given above say
that one should use primary functional coordinates in the group submanifold instead of space-time coordinates. Why?

Coordinates of classical events established by means of the {\it classical electromagnetic field} is based on
the distinguishability, i.e. individualization of material
points. However we loss this possibility with {\it quantum
fields} since we do not have solid scales and ideal clocks acceptable
in the framework of special relativity. The problem
of identification is the root problem even in classical physics and
its recognition gave to Einstein the key to formalization of
the relativistic kinematics and dynamics. Indeed, only assuming the
possibility to detect locally the coincidence of two pointwise
events of a different nature it is possible to build all of the kinematic
scheme and the physical geometry of space-time
\cite{Einstein1,Einstein2}. As such the ``state" of the local clock
gives us local coordinates. In
the classical case the notions of the ``clock" and the ``train" are
intuitively clear and it is assumed that they may be replaced by material points. Furthermore, Einstein especially notes that he
does not discuss the inaccuracy of the simultaneity of two {\it
approximately coincident events} that should be overcame by some
abstraction \cite{Einstein1}. This abstraction is of course the
neglect of finite sizes (and all internal degrees of freedom) of the
both real clock and train. It gives the representation of these
``states" by mathematical points in space-time. Thereby the local
identification of positions of two events is the formal source of the classical
relativistic theory. But generally in the quantum case such identification is
impossible since the space-time coordinates of quantum particles is
state-dependent \cite{NW,A98}. Hence the \emph{quantum identification} of
particles cannot be done in the same manner (as in special relativity) and it requires a physically motivated operational procedure with a corresponding mathematical description. In order to do it some conservation law in the state space
expressing the ``self-identification" should be formulated. The quantum version of the inertial law will be discussed below in connection with this problem
\cite{Le1}.

There were many attempts to build extended models of quantum ``elementary" particles leaving Minkowski space-time structure intact.
I will mention here only Schr\"odinger's attempt to build a stable wave packet as a model of a harmonic oscillator (the first example of coherent state) \cite{Sch2}, Skyrme's soliton solution of sine-Gordon equation \cite{Skyrme}, Dirac's model of the extended electron-muon system \cite{Dirac2}, and the
't Hooft-Polyakov non-singular monopole solution \cite{tHP}. Together with these works intending to build the model of nonlocal quantum particles, it is important to take into account the relativistic quantum non-locality discovered by Newton, Wigner \cite{NW} and Foldy-Wouthuysen \cite{FW} under attempts to reach in fact the opposite target - to find point-like localization of relativistic wave functions.

Here I would like develop essentially different theory of non-local quantum electron where space-time structure arises under objective quantum ``measurement". It is a state dependent gauge field theory based on the intrinsically geometric ``functional" unification of quantum theory and relativity, so-called ``super-relativity" \cite{Le1,Le2,Le3,Le4,Le5,Le6}. Quantum state and geometric classification of their
motion in projective Hilbert space are primordial elements of the new quantum theory. The main assumption is that consistent quantum theory should be based on internal  geometry of quantum state space (in my case it is ``phase space" $CP(N-1)$ diffeomorphic to coset manifold $G/H=SU(N)/S[U(1)\times U(N-1)]$) and that the 4D space-time structure arises only under attempt to ``measure" some quantum dynamical variable, i.e. to establish single value for local dynamical variables (LDV) of the model \cite{Le4}. It means nothing but \emph{the physical ``formulation" of a quantum question is unescapably related to the local Lorentz structure of 4D dynamical space-time.} Thereby the problem of quantum measurement requires the reconstruction of all fundamental notions comprising dynamical space-time structure and geometry of the state space. It turns out that objective quantum measurement is non-distinguishable from space-time structure.

Generally, space-time localization is treated as the ability of coordinate description of an object in classical relativity closely connected with the operational identification  of ``events" \cite{Einstein1}. It is tacitly assumed that all classical objects (frequently represented by material points) are
self-identical and they can not disappear because of the energy-momentum
conservation law. The inertia law of Galileo-Newton ascertains this self-conservation ``externally", i.e. as if one looks on some massive body perfectly isolated from Universe. In such approach only ``mechanical" state of relative motion of the body has been taken into account. Nevertheless, Newton clearly saw some weakness of such approach. His famous example of rotating bucket with water shows that there is an absolute motion since the water takes on a concave shape in any reference frame. Here we are very close to different - ``internal" formulation of the inertia principle and, probably, to understanding  the quantum nature of inertial mass. Namely, the ``absolute motion" of a body should be turned towards not outward, to distant stars, but inward -- to the deformation of the body. This means that external force not only changes the inertial character of its motion:  motion with the constant velocity transforms to accelerated motion, moreover -- the body deforms.. One should have quantum, intrinsic formulation of the inertia law.
One way to establish such formulation may be based on following observation.

Forces not only break the inertial motion of the body
but they generally deform it. This deformation is objective, i.e.
\emph{physical state of body (temporary in a somewhat indefinite sense)} does
not depend on the choice of the inertial reference frame.
I will formulate a quantum inertia law. It paves the way to clarification
the old problem of inertial mass and such ``fictitious" forces as, say,
centrifugal force. Briefly speaking, the inertia and inertial forces
originate not in space-time but in the space of quantum states since they
are generated by deformation of quantum states as a reaction to external interaction.

Up to now the localization problem of quantum systems in the space-time is connected
in fact with the fundamental classical notion of potential energy and force.
Einstein and Schr\"odinger already discussed the inconsistency of usage such
purely classical notions together with the quantum law of motion and the concepts
of ``particle" and ``acceleration" as well (see one of the letter of Einstein
to Schr\"odinger \cite{Ein_Schr} and the article \cite{Ein_de_Broglie}). But
these messages mostly did not get attention by the physical community.

Newton's force is the physical reason for the \emph{absolute} change of the character of motion realized in space-time \emph{acceleration} that serves as geometric
counterpart to force (curvature of the world line in Newtonian space-time
is now non-zero). However there is no adequate geometric notion in quantum
theory since, for example, the notion of trajectory of quantum system was
systematically banned. In some sense the energy of interaction
expressed by $H_{int}$ is analoge of a classical force. Generally,
this interaction leads to the absolute change (deformation) of the quantum
state \cite{Le6} (remember: quantum state is the state of motion
\cite{Dirac}). Motion takes the place in state space modeled
frequently by some Hilbert space. But there is no geometric
counterpart to $H_{int}$ in such functional space. In order to establish the geometric counterpart to $H_{int}$ it is useful
initially to clarify the important question: what is the quantum content
of classical force, if any?

Let me use a small droplet of mercury as simple
example of macroscopic system. The free droplet of mercury is in the state
of the motion ``whether it be of rest, or of moving uniformly forward in a
straight line". Force breaks the inertial character of motion and Galilean
invariance for accelerated system. This statement literally means that physical states of the droplet (its internal degrees of
freedom) in rest and in uniform motion in a straight line are physically
non-distinguishable. The force applied to the droplet deforms its surface
tension, changes its temperature, etc. In fact this external force perturbs
Goldstone's modes supporting the droplet as a macro-system \cite{TFD} and micro-potentials acting on any internal quantum particle, say, electrons inside of the droplet. It means that quantum states and their deformations may serve as a
``detector" of the ``external force" action on the droplet. Therefore it is
reasonable to use quantum state deformations instead of classical acceleration,
since generally acceleration depends on mass, charge, etc., that is
impossible establish a pure space-time invariant (geometric) counterpart
of a classical force independent on material body. Only a classical gravitation
force may be geometrized assuming the gravitation field may be replaced locally by
an accelerated reference frame since in general relativity the gravitation
and reference frame are locally non-distinguishable.
There is, however, a more serious reason why space-time acceleration is not a
so good counterpart of the force.

The physical state of the droplet freely falling in the gravitation field
of a star is non-distinguishable from the physical state of the droplet in
an remote from stars area. Therefore macroscopic space-time acceleration cannot serve as a discriminator of physical state of body \cite{Le5,Le7}. Thus, instead of choosing, say, the system of distant stars as an ``outer" absolute reference frame
\cite{Einstein2} the deformation of quantum state of some particle of the droplet may be used. It means that the deformation of quantum motion in quantum state space serves as an ``internal detector" for ``accelerated" space-time motion.

I have assumed that same approach may be applied to single quantum electron whose model is a dynamical process in the state space of its spin/charge degrees of freedom. Then the deformations of quantum motion generated by the coset action in the quantum state space will be used as an internal counterpart of a self-interacting electron in dynamical space-time. It means nothing but in the developing theory \emph{a distance between quantum states in the state space should replace a distance between ``bodies" in space-time as the primary geometric notion}.

\section{Coset transformations vs F.-W.}
Dirac clearly understood that the electron is a non-local particle since it has internal structure \cite{Dirac2,Dirac3}.
Nevertheless, he successfully formulated linear relativistic wave equation for point-like particle. The next step should be achieved in the procedure of second quantization of the bi-spinor amplitudes in order to take into account self-interaction, creation of pairs, etc. However higher orders of perturbation being formally applied to equations of motion arose under this ``stiff" method of quantization lead to divergences \cite{Dirac4} even in the case of QED with small coupling fine structure constant $\alpha=e^2/\hbar c \approx 0.007$.

Trying to study non-local structure of electron, I avoid to use the second quantization \cite{Dirac1} using instead the smooth vector fields playing the role of LDV's (local dynamical variables) of Dirac's electron \cite{Le1,Le4}. The main aim is to get non-linear wave
equation expressing the conservation law of relativistic energy-momentum operator and to study its lump (soliton-like) solution for the ``field shell" associated with the surrounding field of a single electron. This equation should have solutions similar to well known 't Hooft-Polyakov regular monopole solution with finite energy \cite{tHP} but without additional Higgs fields. It should be proved (it is not done yet) that the ``field shell" \emph{integrally} contains all processes treated in the standard QFT as vacuum polarization, etc.

The extremal of the least action principle, say, solution of the ordinary Dirac equation is the plane wave ``modulated" by a bi-spinor
\begin{eqnarray}
|\Psi(x)>=\left(                                                               \begin{array}{cc}
\psi_1  \\
\psi_2 \\
\psi_3 \\
\psi_4 \\
\end{array}
\right) \exp{\frac{-i}{\hbar}P_{\mu}x^{\mu}}.
\end{eqnarray}
The plane wave is the improper state of the quantum action with an arbitrary mass and momentum connected only by the ``on-shell" dispersion law.
If we try to find some invariant physics of electron underlying the states generated by the flexible ``setup" described by the LDV's, one should vary the conservation law of quantum numbers.

The relativistic Klein-Gordon and Dirac wave equations are based on the classical mass-shell relation $p^{\mu}p_{\mu}-m^2c^2=0$. The last one is so restrictive
that most consequences of the Dirac equations almost literally coincide with
classical equations of motions \cite{Fock}. However further quantum corrections generated by second quantization destroy in fact this idealized picture: there is a diffusion of the mass-shell due to space-like self-interaction effects.
On the other hand the Foldy-Wouthuysen (F.-W.) unitary transformations
\begin{eqnarray}
U=e^{iS}= \cos |\mathbf{P}|\theta + \frac{\beta \vec{\alpha}}{|\mathbf{P}|}\sin| \mathbf{P}|\theta, \quad U \in SU(4)
\end{eqnarray}
\cite{FW}
reveal already the non-local nature of electron wave function without any references to second quantization. It is interesting therefore to get more general relativistic non-linear wave equation which has non-singular localizable solution associated with quantum particles.

F.-W. unitary transformations were invoked to diagonalize Dirac's Hamiltonian in order to separate bi-spinor components with positive and negative energies. Generally, diagonalization is exactly possible in the case of free electron and approximately for an electron in external fields. This transformation is non-local since it effectively delocalises pointlike electron in a spatial vicinity with the radius $\delta r \approx \frac{\hbar}{mc}$. The mass of electron $m$ is a free parameter of the model.

I should note two peculiarities of the F.-W. transformations.
First, these transformations intended to separate ``large" and ``small" components with positive and negative energies. This ``large/small" classification is  neither unitary nor scale invariant. The approximate diagonalization chosen by F.-W. is based on the iteration scheme of the Hamiltonian corrections in moving frame in one ``slowness" parameter $\frac{\hbar}{mc}$ that is scale non-invariant. However the scale-invariant local projective coordinates $(\pi^1, \pi^2, \pi^3)$  in $CP(3)$ (see below) may be used in order to build local reference frame in which the operator of energy-momentum of the self-interacting electron is ``instantly diagonal".

Second, partially separated quantum degrees of freedom (spin and charge) and the space-time coordinates lead (under F.-W. transformations) to delocalization of electron as we will discuss below. The analysis of the ``delocalization" leads however to some progress in understanding quantum dynamics of electron.

In order to understand the reason of delocalization arose as a result of diagonalization one need to take into account the geometry of $SU(4)$ group and the Cartan decomposition of $AlgSU(4)$ \cite{KN,Le1,Le2,Le3}. First of all we shall note that the matrices
\begin{eqnarray}
\hat{\gamma}_0= \left(
\matrix{1&0&0&0 \cr
0&1&0&0 \cr
0&0&-1&0 \cr
0&0&0&-1} \right), \quad
\hat{\gamma}_1= \left(
\matrix{0&0&0&-i \cr
0&0&-i&0 \cr
0&i&0&0 \cr
i&0&0&0} \right),\cr
\hat{\gamma}_2= \left(
\matrix{0&0&0&-1 \cr
0&0&1&0 \cr
0&1&0&0 \cr
-1&0&0&0} \right),
\hat{\gamma}_3= \left(
\matrix{0&0&-i&0 \cr
0&0&0&i \cr
i&0&0&0 \cr
0&-i&0&0} \right),
\end{eqnarray}
originally introduced by Dirac \cite{Dirac3}  may be represented as linear combinations of ``standard" $SU(4)$ $\lambda$-generators \cite{Close}
\begin{eqnarray}
\hat{\gamma}_0&=&\hat{\lambda}_3+\frac{1}{3}[\sqrt{3}\hat{\lambda}_8-
\sqrt{6}\hat{\lambda}_{15}],\cr
\hat{\gamma}_1&=&\hat{\lambda}_2+\hat{\lambda}_{14},\cr
\hat{\gamma}_2&=&\hat{\lambda}_1-\hat{\lambda}_{13},\cr
\hat{\gamma}_3&=&-\hat{\lambda}_5+\hat{\lambda}_{12}.
\end{eqnarray}
Since any state $|S>$ has the isotropy group
$H=U(1)\times U(N)$, only the coset transformations $G/H=SU(N)/S[U(1)
\times U(N-1)]=CP(N-1)$ effectively act in $C^N$. One should remember,
however, that the concrete representation of hermitian matrices
belonging to subsets $h$ or $b$ (as defined below) depends on a priori chosen vector (all ``standart" classification of the traceless matrices of Pauli, Gell-Mann, etc., is based on the vector $(1,0,0,...,0)^T$).   \emph{The Cartan's decomposition of the algebra $AlgSU(N)$ is unitary invariant and I will use it instead of Foldy-Wouthuysen decomposition in ``even" and ``odd" components.}

According to Cartan's classification of $SU(4)$ generators, there are two types of
generators $\hat{\gamma}$ relative the state vector
\begin{eqnarray}
|S(x)>=\left(
\begin{array}{cc}
\psi_1  \\
0 \\
0 \\
0 \\
\end{array}
\right) \exp{\frac{-i}{\hbar}P_{\mu}x^{\mu}}:
\end{eqnarray}                                                                        $\hat{\gamma}_0$ is generator of the isotropy group $H=U(1)\times U(3)$ of the
$|S(x)>$ leaving it intact, whereas $\hat{\gamma}_1, \hat{\gamma}_2, \hat{\gamma}_3$
belong to coset transformations $G/H=SU(4)/S[U(1)\times U(3)=CP(3)$ that deform the
chosen state. The algebra of generators $AlgSU(N)$ is $Z_2$-graded in respect with
following properties of the commutation relations.
$[h_{|S>},h_{|S>}] \subseteq h_{|S>}, [b_{|S>},b_{|S>}]
\subseteq h_{|S>}, [b_{|S>},h_{|S>}] \subseteq b_{|S>}$. One may easy check that for
example $\hat{\gamma}_1, \hat{\gamma}_2 \subseteq b_{|S>}$ and
\begin{eqnarray}
\hat{\gamma}_1 \hat{\gamma}_2 -\hat{\gamma}_2 \hat{\gamma}_1
=2i[\frac{1}{3}(\sqrt{3}\hat{\lambda}_8-                                              \sqrt{6}\hat{\lambda}_{15})-\hat{\lambda}_3]\subseteq h_{|S>}.
\end{eqnarray}
Physically it is important to use the Cartan decomposition of unitary group in
 respect with initially chosen state vector $|S>$.
 Therefore the parametrization of these decomposition is state-dependent
 $[h_{|S>},h_{|S>}] \subseteq h_{|S>}, [b_{|S>},b_{|S>}]
 \subseteq h_{|S>}, [b_{|S>},h_{|S>}] \subseteq b_{|S>}$ \cite{Le1,Le2,Le3}.                  It means that physically it is interesting not abstract unitary group relations              but realization of the unitary transformations resulting in motion of the pure               quantum states represented by rays in projective Hilbert space. Therefore the                ray representation of $SU(N)$ in $C^N$, in particular, the embedding
 of $H$ and $G/H$ in $G$, is a state-dependent parametrization.
The diagonalization of the Dirac's Hamiltonian is the annihilation of the coset part of the initial Hamiltonian acting on the bi-spinor in $C^4$. Notice, that the ``modulated" plane wave (1) belongs to functional space $\mathcal{H}=C^4 \otimes C^{\infty}$ which is the tensor product. But unitary operator (2) capable diagonalize only matrix part acting on the bi-spinor components in $C^4$ and it does not commute with operator of the coordinate that leads to non-singular function
\begin{eqnarray}
<\mathbf{r}|\hat{U}|\mathbf{r'}>=(2\pi)^{-3}\int [\sqrt{\frac{m+E_p}{2E_p}}
+\hat{\beta}\frac{\mathbf{\hat{\alpha}}\mathbf{p}}{\sqrt{2E_p(m+E_p)}}]e^{i\mathbf{p}
(\mathbf{r}-\mathbf{r'})}d\mathbf{p}
\end{eqnarray}
\cite{Messia} in contrast with the $\delta$-function.
In other words F.-W. transformations could not leave intact the plane waves thus they create the superposition of improper states in the Hilbert space denoted here as $C^{\infty}$. Therefore according to general classification of quantum motion \cite{Le1} the coset transformation is a quantum analog of force giving deformation of quantum state. This leads to delocalization of electron and I try to develop this result attempting to derive non-linear relativistic quantum field equations in the spirit of approach proposed a few years before \cite{Le1,Le3,Le4}.
The main aspiration is to find new non-linear wave equation for energy-momentum of electron moving in dynamical space-time.

The local projective coordinates coordinates of eigenstate
\begin{equation}
\pi^i_{(j)}=\cases{\frac{\psi^i}{\psi^j},&if $ 1 \leq i < j$ \cr
\frac{\psi^{i+1}}{\psi^j}&if $j \leq i < 4$}
\end{equation}\label{11}
in the map $U_j:\{|\Psi>,|\psi^j| \neq 0 \},1 \leq j \leq 4$
of free electron in $CP(3)$ may be derived from ordinary homogeneous system of eigen-problem
\begin{eqnarray}
mc^2 \psi_1+c(p_x-ip_y)\psi_4+cp_z \psi_3=E\psi_1 \cr
mc^2 \psi_2+c(p_x+ip_y)\psi_3-cp_z \psi_4=E\psi_2 \cr
-mc^2 \psi_3+c(p_x-ip_y)\psi_2+cp_z \psi_1=E\psi_3 \cr
-mc^2 \psi_4+c(p_x+ip_y)\psi_1-cp_z \psi_2=E\psi_4 .
\end{eqnarray}\label{12}

It is easy to see \cite{Le3} that under transition from the system of homogeneous
equations to the reduced system of non-homogeneous equations (the first equation was omitted)
\begin{eqnarray}
(-E+m c^2) \pi^1+c(P_x+iP_y) \pi^2-c P_z \pi^3 &=& 0 \cr
c(P_x-iP_y) \pi^1-(E+m c^2) \pi^2 &=& cP_z \cr
-cP_z \pi^1-(E+m c^2) \pi^3 &=& c (P_x+iP_y),
\end{eqnarray}\label{13}
one has the single-value solution for eigen-ray
\begin{eqnarray}\label{13}
\pi^1=0, \quad
\pi^2=\frac{-cp_z}{E+mc^2} \quad
\pi^3=\frac{-c(p_x+ip_y)}{E+mc^2},
\end{eqnarray}
in the map $U_1:\{\psi_1 \neq 0\}$ for $E=\sqrt{m^2c^4+c^2p^2}+\delta$
and the non-homogeneous equations (the forth equation was omitted)
\begin{eqnarray}
(E-m c^2) \pi^1-c P_z \pi^3 &=& c(P_x-iP_y) \cr
(E-m c^2) \pi^2-c (P_x+iP_y) \pi^3 &=& -cP_z \cr
cP_z \pi^1+ c (P_x-iP_y) \pi^2 -(E+m c^2) \pi^3 &=& 0
\end{eqnarray}\label{13}
in the map $U_4:\{\psi_4 \neq 0\}$ for $E=-\sqrt{m^2c^4+c^2p^2}-\delta$
with the solution
\begin{eqnarray}\label{13}
\pi^1=\frac{c(p_x-ip_y)}{E-mc^2}, \quad
\pi^2=\frac{-cp_z}{E-mc^2} \quad
\pi^3=0.
\end{eqnarray}
It is possible only if the determinant of the reduced system
$D=(E^2-m^2c^4-c^2p^2)^2 $ is not vanished. It is naturally to use these scale-invariant functional variables $(\pi^1,\pi^2,\pi^3)$ in order to establish relation between spin-charge degrees of freedom
and energy-momentum distribution of electron in dynamical space-time (DST) since
the ``off-shell" condition $D=(E^2-m^2c^4-c^2p^2)^2 \neq 0$ opens the way
for self-interaction. New dispersion law will be established
due to formulation of the conservation law of quantum energy-momentum.
In local coordinates (representation) the improper states like plane waves
are simply deleted. It means that trivial free motion of whole quantum
setup in local homogeneous space-time is removed.

\section{Energy-momentum operator as a tangent vector to $CP(3)$}
Since it is impossible to find the representation capable exactly to diagonalize Hamiltonian with help of global non-Abelian actions of $SU(4)$ in dynamical situation and because even in the case of free electron the diagonalization is achievable only in one sub-space $C^4$ of full state space, let me reformulate the problem as follows.

I will work with Dirac's operator of energy-momentum
\begin{equation}\label{}
 \hat{\gamma}^{\mu}p_{\mu} =i\hbar \hat{\gamma}^{\mu}\frac{\partial}{\partial x^{\mu}}
\end{equation}
instead of the Hamiltonian. This combined operator acts in the direct product $S=C^4 \times H_D$, where $H_D$ means a Hilbert space of differentiable functions.
Such splitting seems to be artificial and I try to find a more flexible construction
of energy-momentum operator. Lets apply to this operator the similarity transformation (transition to ``moving frame" freezing the action of the differentiation in space-time coordinates) with help of the canonical unitary operator. In the case of pseudo-euclidian coordinates $x^{\mu}$ it is possible to use simply the ``plane wave" $U_{gauge}=exp(-\frac{i}{\hbar}P_{\mu}x^{\mu})$. But if one uses, say, spherical coordinates, one needs to use non-Abelian gauge transformations of $SU(4)$ in order to convert the operator into a matrix with functional elements. I will discuss this in a separate paper.

The most delicate point of the construction is as follows.
Energy-momentum variation evoked by the internal dynamical structure of electron is independent of the global space-time transformation (being applied to electron's ``center of mass") but nevertheless it should be reflected in space-time motion of the concentrated ``field-shell" \cite{LeHor}. This may be treated as a result of the back-reaction from quantum dynamics of spin and charge degrees of freedom during ``metabolic time" controlling the motion in the state space $CP(3)$. ``Observable" soliton-like dynamics arises under the lift in the tangent fibre bundle from the base manifold $CP(3)$ into the state dependent ``dynamical space-time" (DST) \cite{Le2} that will be discussed below.

I assume that $P_{\mu}=P_{\mu}(\tau)$ is the function of the ``proper time" $\tau$ and consequently the function of state-dependent dynamical space-time coordinates that will be introduced only on the stage of ``quantum measurement" \cite{Le3}. Then one has the matrix
\begin{equation}\label{}
 U_{gauge}^{-1} \hat{\gamma}^{\mu} p_{\mu} U_{gauge} =  \hat{\gamma}^{\mu} P_{\mu}(\tau)
\end{equation}
with functional coefficients, not operator-valued.

Infinitesimal energy-momentum  variations  evoked by interaction
charge-spin degrees of freedom (implicit in $\hat{\gamma}^{\mu}$ ) that may be
expressed in terms of local coordinates $\pi^i$ since there is a
diffeomorphism between the space of the rays $CP(3)$ and the $SU(4)$
group sub-manifold of the coset transformations
$G/H=SU(4)/S[U(1) \times U(3)]=CP(3)$ and the isotropy group $H=U(1) \times U(3)$
of some state vector. It will be expressed by the
coefficient functions of combinations of the $SU(4)$ generators $\hat{\gamma}_{\mu}$
of unitary transformations that will be defined by an equation arising
under infinitesimal variation of the energy-momentum
\begin{equation}
\Phi_{\mu}^i(\gamma_{\mu}) = \lim_{\epsilon \to 0} \epsilon^{-1}
\biggl\{\frac{[\exp(i\epsilon \hat{\gamma}_{\mu})]_m^i \psi^m}{[\exp(i
\epsilon \hat{\gamma}_{\mu})]_m^j \psi^m }-\frac{\psi^i}{\psi^j} \biggr\}=
\lim_{\epsilon \to 0} \epsilon^{-1} \{ \pi^i(\epsilon
\hat{\gamma}_{\mu}) -\pi^i \},
\end{equation}\label{14}
arose in a nonlinear local realization of $SU(4)$ \cite{Le1}. Here
$\psi^m, 1\leq m \leq 4$ are ordinary bi-spinor amplitudes. I calculated the twelve coefficient functions  $\Phi_{\mu}^i(\gamma_{\mu})$ in the map $U_1:\{\psi_1 \neq 0\}$:
\begin{eqnarray}
\Phi_{0}^1(\gamma_{0})&=&0, \quad \Phi_{0}^2(\gamma_{0})=-2i\pi^2,
\quad \Phi_{0}^3(\gamma_{0})=-2i\pi^3; \cr
\Phi_{1}^1(\gamma_{1})&=&\pi^2 -\pi^1 \pi^3,
\quad \Phi_{1}^2(\gamma_{1})=-\pi^1 -\pi^2 \pi^3,
\quad \Phi_{1}^3(\gamma_{1})=-1 -(\pi^3)^2; \cr
\Phi_{2}^1(\gamma_{2})&=&i(\pi^2 +\pi^1 \pi^3),
\quad \Phi_{2}^2(\gamma_{2})=i(\pi^1 +\pi^2 \pi^3),
\quad \Phi_{2}^3(\gamma_{2})=i(-1 +(\pi^3)^2); \cr
\Phi_{3}^1(\gamma_{3})&=&-\pi^3 -\pi^1 \pi^2,
\quad \Phi_{3}^2(\gamma_{3})=-1 -(\pi^2)^2,
\Phi_{3}^3(\gamma_{3})=\pi^1 -\pi^2 \pi^3.
\end{eqnarray}\label{15}

Now I will define the $\Gamma$-vector field
\begin{equation}
\vec{\Gamma}_{\mu}=\Phi_{\mu}^i(\pi^1,\pi^2,\pi^3)\frac{\partial}{\partial \pi^i}
\end{equation}\label{16}
and then the energy-momentum operator will be defined as the \emph{functional
vector field}
\begin{equation}\label{17}
P^{\mu}\vec{\Gamma}_{\mu}\Psi(\pi^1,\pi^2,\pi^3)
= P^{\mu}\Phi_{\mu}^i(\pi^1,\pi^2,\pi^3)
\frac{\partial}{\partial \pi^i}\Psi(\pi^1,\pi^2,\pi^3) + c.c.
\end{equation}
acting on the ``total wave function",
where the ordinary 4-momentum $P^{\mu}=(\frac{E}{c}-\frac{e}{c}\phi,\vec{P} -
\frac{e}{c} \vec{A})=(\frac{\hbar \omega}{c}-\frac{e}{c}\phi,\vec{\hbar k} -
\frac{e}{c} \vec{A})$ (not operator-valued) should be identified with
the solution of quasi-linear ``field-shell" PDE's  for the contravariant
components of the energy-momentum tangent vector field in $CP(3)$
\begin{equation}\label{18}
P^i(x,\pi)=P^{\mu}(x)\Phi_{\mu}^i(\pi^1,\pi^2,\pi^3).
\end{equation}
In some sense the ``total wave function"
$\Psi = \Psi(\pi^1,\pi^2,\pi^3)$ of local coordinates
$\pi^i$  is similar to the non-bilinear function
$a(\psi,\psi^*)$ of Weinberg \cite{Weinberg} but without requirement of homogeneous
of degree one in $\pi$ and $\pi^*$.
One sees that infinitesimal variation of energy-momentum is represented by
the operator of partial differentiation in complex local coordinates $\pi^i$
with corresponding coefficient functions $\Phi_{\mu}^i(\pi^1,\pi^2,\pi^3)$.
Then the single-component ``total wave function"
$\Psi(\pi^1,\pi^2,\pi^3)$ should be
studied in the framework of new PDE instead of two-component approximation
due to Foldy-Wouthuysen unitary transformations. There are of course four such
functions $\Psi(\pi^1_{(1)},\pi^2_{(1)},\pi^3_{(1)})$, $\Psi(\pi^1_{(2)},
\pi^2_{(2)},\pi^3_{(2)})$,
$\Psi(\pi^1_{(3)},\pi^2_{(3)},
\pi^3_{(3)}), \Psi(\pi^1_{(4)},\pi^2_{(4)},
\pi^3_{(4)})$
- one function in each local map.

\section{Eigen-dynamics and local dynamical variables}
The standard QM tells us
what is the spectrum of quantum dynamical variables but it is silent about dynamics
of morphogenesis of stationary quantum states (since they are states of motion).
There were however essential efforts intended to clarify the process of
``state vector reduction"  in the framework of modified Schr\"odinger
equation \cite{Hughston,AdHor} with help stochastic additional terms
without evident physical ``mechanism" at a deeper level. I use state-dependent
projective representation of the local dynamical variables (LDV) of
internal degrees of freedom associated with the generators of $SU(N)$ \cite{Le4}.

Since the system of eigen-vectors belonging to degenerate eigenvalues is defined up to unitary transformations, the approximate calculation of eigenvalues and corresponding eigen-state vectors in the conditions of degeneration is natural place for the application of geometry of unitary group. For example, the solution of the problem of small denominators arising in the framework of perturbation theory is based in fact on the geometry of $CP(1)$, see for example \cite{Fock}.

Pseudo-electric and pseudo-magnetic fields arose as gauge fields with singular potentials at the degeneracy points of the Hamiltonian spectrum \cite{Berry1}.
The structure of degeneration is unstable to a relative small perturbation of the Hamiltonian and hence could not serve as a source of real electromagnetic potentials.

I would like to study the nature of affine unitary gauge fields arising under breakdown (reconstruction) of global $G=SU(4)$ symmetry to the local gauge group $H=S[U(1)\times U(3)]$ acting with state-dependent generators on ``phase space" $CP(3)$.

The quantum mechanics assumes the priority of the Hamiltonian given
by some classical model which henceforth should be ``quantized". It
is known that this procedure is ambiguous. In order to avoid the
ambiguity, I intend to use a {\it quantum state} itself and the
invariant conditions of its conservation and perturbation. These
invariant conditions are rooted into the global geometry of the
dynamical group manifold. Namely, the geometry of $G=SU(N)$, the
isotropy group $H=U(1)\times U(N-1)$ of the pure quantum state, and
the coset $G/H=SU(N)/S[U(1)\times U(N-1)]$ geometry, play an essential
role in the quantum state evolution \cite{Le4}. The stationary states
i.e. the states of motion with the least action may be
treated as {\it initial conditions} for GCS evolution. Particulary
they may represent a local minimum of energy (local vacuum).

Now I will introduce the local dynamical variables (LDV's) correspond
to the internal $SU(N)$ group symmetry and its breakdown. They should be
expressed now in terms of the local coordinates $\pi^k$. Thereby they
will live in geometry of $CP(N-1)$ with the Fubini-Study metric
\begin{equation}
G_{ik^*} = [(1+ \sum |\pi^s|^2) \delta_{ik}- \pi^{i^*} \pi^k](1+
\sum |\pi^s|^2)^{-2}.
\end{equation}\label{2}
Hence the internal dynamical
variables and their norms should be state-dependent, i.e. local in
the state space \cite{Le4}. These local dynamical variables realize
a non-linear representation of the unitary global $SU(N)$ group in
the Hilbert state space $C^N$. Namely, $N^2-1$ generators of $G =
SU(N)$ may be divided in accordance with the Cartan decomposition:
$[B,B] \in H, [B,H] \in B, [B,B] \in H$. The $(N-1)^2$ generators
\begin{eqnarray}
\Phi_h^i \frac{\partial}{\partial \pi^i}+c.c. \in H,\quad 1 \le h
\le (N-1)^2
\end{eqnarray}\label{3}
of the isotropy group $H = U(1)\times U(N-1)$ of the ray (Cartan
sub-algebra) and $2(N-1)$ generators
\begin{eqnarray}
\Phi_b^i \frac{\partial}{\partial \pi^i} + c.c. \in B, \quad 1 \le b
\le 2(N-1)
\end{eqnarray}\label{4}
are the coset $G/H = SU(N)/S[U(1) \times U(N-1)]$ generators
realizing the breakdown of the $G = SU(N)$ symmetry of the GCS.
Furthermore, the $(N-1)^2$ generators of the Cartan sub-algebra may
be divided into the two sets of operators: $1 \le c \le N-1$ ($N-1$
is the rank of $Alg SU(N)$) Abelian operators, and $1 \le q \le
(N-1)(N-2)$ non-Abelian operators corresponding to the
non-commutative part of the Cartan sub-algebra of the isotropy
(gauge) group. Here $\Phi^i_{\sigma}, \quad 1 \le \sigma \le N^2-1 $
are the coefficient functions of the generators of the non-linear
$SU(N)$ realization. They give the infinitesimal shift of the
$i$-component of the coherent state driven by the $\sigma$-component
of the unitary  field $\exp(i\epsilon \lambda_{\sigma})$ rotating by the
generators of $Alg SU(N)$ and they are defined as follows:
\begin{equation}
\Phi_{\sigma}^i = \lim_{\epsilon \to 0} \epsilon^{-1}
\biggl\{\frac{[\exp(i\epsilon \lambda_{\sigma})]_m^i \psi^m}{[\exp(i
\epsilon \lambda_{\sigma})]_m^j \psi^m }-\frac{\psi^i}{\psi^j} \biggr\}=
\lim_{\epsilon \to 0} \epsilon^{-1} \{ \pi^i(\epsilon
\lambda_{\sigma}) -\pi^i \},
\end{equation}\label{5}
\cite{Le2,Le4}. Then each of the $N^2-1$ generators
may be represented by vector fields comprised by the coefficient
functions $\Phi_{\sigma}^i$ contracted with corresponding partial derivatives
$\frac{\partial }{\partial \pi^i} = \frac{1}{2}
(\frac{\partial }{\partial \Re{\pi^i}} - i \frac{\partial }{\partial
\Im{\pi^i}})$ and $\frac{\partial }{\partial \pi^{*i}} = \frac{1}{2}
(\frac{\partial }{\partial \Re{\pi^i}} + i \frac{\partial }{\partial
\Im{\pi^i}})$.
Now one may define the ``flexible quantum setup" as the local reference
frame in tangent space $TCP(N-1)$ to the projective Hilbert state space $CP(N-1)$
\begin{eqnarray}\label{6}
\{e_{\sigma}, e_{\sigma}^*\} =
\{\Phi^i_{\sigma}\frac{\partial}{\partial \pi^i},
\Phi^{*i}_{\sigma}\frac{\partial}{\partial \pi^{*i}}\}.
\end{eqnarray}
LDV's may be built from these vector fields corresponding to $SU(N)$ generators
acting on quantum state.

I would like to give some general notes concerning eigen-dynamics:

The anholonomy of the wave function arose due to slowly variable
environment was widely discussed by Berry and many other authors in the
framework of so-called geometric phases \cite{Berry1}. It is clear that
now we deal with different problem: \emph{Berry made accent on variation
of wave function during cycle motion whereas for us interesting the quantum invariants of infinitesimal variation of the quantum setup}.

The geometric phase is an intrinsic property of the family
of eigenstates. There are in fact a set of  local dynamical variables (LDV)
that like the geometric phase
intrinsically depends on eigenstates. For us will be interesting only the set
comprising vector field
$\xi^k(\pi^1,...,\pi^{N-1}): CP(N-1)\rightarrow \mathcal{C}$
in local coordinates $\pi^i_{(j)}$. In view of future discussion of
F.-W. ``instant" transformations it is useful
to compare \emph{velocity} of variation of the Berry's phase
\begin{equation}
\dot{\gamma}_n(t) = -\textbf{A}_n(\textbf{R})\dot{\textbf{R}},
\end{equation}\label{7}
where $\textbf{A}_n(\textbf{R})=\Im<n(\textbf{R})| \nabla_{\textbf{R}} n(\textbf{R})>$
with the affine parallel transport of the vector field $\xi^k(\pi^1,...,\pi^{N-1})$
given by the equations
\begin{equation}
\frac{d \xi^i}{d\tau}=-\Gamma^i_{kl}\xi^k \frac{d \pi^l}{d\tau}.
\end{equation}\label{8}
The parallel transport of Berry is similar to the affine parallel transport but
the last one is fundamental in respect of gauge and scale-invariant conservation
law agrees with Fubini-Study ``quantum metric tensor" $G_{ik^*}$ in the base manifold
$CP(N-1)$.
The affine gauge field given by connection
\begin{eqnarray}
\Gamma^i_{mn} = \frac{1}{2}G^{ip^*} (\frac{\partial
G_{mp^*}}{\partial \pi^n} + \frac{\partial G_{p^*n}}{\partial
\pi^m}) = -  \frac{\delta^i_m \pi^{n^*} + \delta^i_n \pi^{m^*}}{1+
\sum |\pi^s|^2} \label{Gamma}
\end{eqnarray}\label{9}
is of course more close to the Wilczek-Zee
non-Abelian gauge fields $A_{ab}=(\psi_b|\dot{\psi}_a)$ \cite{WZ}.
Time-dependent choice of the functional basis $\psi'(t)=\Omega(t)\psi(t)$
leads to proper gauge transformation of
$A'(t)=\Omega A(t)\Omega^{-1}+\dot{\Omega}\Omega^{-1}$.
One has formally very similar transformation law for the connection form
$\Gamma^i_k = \Gamma^i_{kl} d \pi^l$ in $CP(N-1)$:
\begin{eqnarray}
\Gamma'^i_k = \Lambda^i_m\Gamma^m_j \Lambda^{-1j}_k+d\Lambda^i_s \Lambda^{-1s}_k,
\end{eqnarray}\label{10}
however there are serious mathematical and physical differences between global
character of gauge transformations
of $A(t)$ in linear state space $C^N$ and local (state-dependent) gauge
transformations of $\Gamma ^i_k = \Gamma^i_{kl} d \pi^l$ in $CP(N-1)$.
Namely, we use from the mathematical point of view the Cartan's method of
moving reference
frame as a ``flexible quantum setup", i.e. reference frame whose motion refers to
itself with infinitesimally close coordinates. This releases us from the necessity to use so-called ``second particle" \cite{AA88,Le2,Le3,Le4} as an external
reference frame. The physical sense of the difference is that we do not
have in our problem environmental ``external" field in which the quantum
system is ``immersed".
On the contrary, we study gauge transformations generated ``from
nowhere", i.e. from pure internal quantum dynamics in the geometry of
state space $CP(N-1)$  and its emergence in the dynamical space-time (DST).
This approach is based on the conception of the quantum inertia \cite{Le1} since
the affine parallel transport of energy-momentum vector field in $CP(N-1)$ expresses
self-conservation of, say, electron.

\section{State space and dynamical space-time}
How internal quantum degrees of freedom of electron may be mapped in dynamical space-time? Can to lift it from the base $CP(3)$ into the tangent fibre bundle?
If we assume that internal dynamics is represented by LDV like an energy momentum vector field $\vec{P}$ then it is natural to think that the process of measurement of the \emph{value} of the LDV should be somehow connected with this mapping. Being geometrically formulated, i.e. in an invariant manner, this process will be objective.
In such case the ``measurement" means only the process of some ``projection" in an attempt to find a single value of physical dynamical variable.
It means that objective quantum measurement is in fact an embedding of quantum dynamics in space-time. But how we should to do it in an invariant, geometric manner? And what is the space-time itself without such measurement?

I use an approach that I called  ``inverse representation"
\cite{Le1,Le2,Le3,Le4,Le5} where finite dimension $N$ state vector of
internal degrees of freedom and velocities of its variation are induced
by the unitary group $SU(N)$ and its actions should be realized by the
lump motion in dynamical space-time (DST).
It means that the space-time degrees of freedom and space-time geometry
should be derived in order to describe energy-momentum distribution in the
``field shell" wrapping internal degrees of freedom. It is useful to
refer here on an analogy with MRI approach to visualization of nuclear spins.

The MRI method may be shortly formulated as follows: ``Many scientists
were taught that you can not image objects smaller than the wavelength of the
energy being used to image. MRI gets around this limitation by producing images
based on spatial variations in the phase and frequency of the radio frequency
energy being absorbed and emitted by the imaged object" \cite{MRI}.
My approach is similar to this method but, say, going in opposite direction:
space-time localization of quantum  particle, say, electron, arises due to
infinitesimal $SU(4)$ variations of its quantum state in projective Hilbert
state space $CP(3)$ that generate self-interacting non-Abelian affine gauge field
agrees with Fubuni-Study metric.
The ``field-shell" of electron's energy-momentum obeys to quasi-linear PDE's
whose soliton-like solution in dynamical space-time is similar to a voxel in
time-dependent slice used instead of a pixel in two-dimension MRI picture.
These PDE's express the conservation law of energy-momentum field. They are
self-consistent with the system of algebraic equations expressing  a result
of the ``comparison" of the infinitesimal parallel transported LDV that may
be encoded by a qubit spinor.  Two infinitesimally
close qubit spinors being connected by infinitesimal $SL(2,C)$ transformation
associated with Lorentz frame transformation generated by ``quantum boost"
and ``quantum rotation". These boosts and rotations are associated with
electromagnetic-like fields. It means that the physical nature of variation of
quantum state being initially taken in the kinematical manner (formal transition
from one inertial frame to another) looks as EM-like fields in the
dynamical variation. Dynamical attachment of the Lorentz frame is analog of the
transition to rotation frame in MRI but the construction of the frame is
of course different.

Absolute values (reference frame independent) of space and time intervals lose their sense in the framework of relativity and only the space-time interval has invariant sense. The distance between quantum states is the main quantum invariant that replaces the space-time interval between two events in the present approach. Thereby the geometry of $CP(N-1)$ defines quantum dynamics of local dynamical variables.
Self-conservation of the electron may be expressed through an affine parallel transported energy-momentum field $P^i$ and the ``measurement procedure" is necessary for recovering the distribution of energy-momentum in dynamical space-time.
It is a place to compare the ``standard" spontaneous generating mass \cite{Guiedry} by the Higss mechanism and the self-interacting construction.

The qubit spinor
\begin{eqnarray}
\eta=\left(
  \begin{array}{cc}
    \eta^0  \cr
    \eta^1
  \end{array}
\right)
\end{eqnarray}
has been built in order to form ``yes/no" quantum question
in respect of separation of the coset action  $G/H=SU(N)/S[U(1)
\times U(N-1)]=CP(N-1)$ on the quantum state $|S>$ and
its isotropy group action $H=U(1)\times U(N)$ used for the ``measurement" of LDV $D^i$. It is assumed that this spinor should replace a complex $SU(2)$ doublet of Lorentz scalar fields
\begin{eqnarray}
\phi=\frac{1}{\sqrt{2}}\left(
  \begin{array}{cc}
    \phi_1+i\phi_2  \cr
    \phi_3+i\phi_4
  \end{array}
\right)
\end{eqnarray}
involved in the Lagrangian density
\begin{eqnarray}
\mathcal{L}_s=(D^{\mu} \phi)^\dag(D_{\mu} \phi)-V(\phi^\dag\phi)=(D^{\mu} \phi)^\dag(D_{\mu} \phi)
-\mu^2 \phi^\dag\phi-\lambda(\phi^\dag\phi)^2,
\end{eqnarray}
where $D_{\mu}= \partial_{\mu}+\frac{ig'}{2}a_{\mu}y+\frac{ig}{2}\vec{\tau}\vec{b}_{\mu}$ is so-called ``covariant derivative" \cite{Guiedry}.

It was not intended here to establish the connection of this model with the Standard Model. This will be a next step of the investigation. I discuss in this work the  eigen-dynamics of relativistic self-interacting electron due to the breakdown of $SU(4)$ to $H=U(1) \times U(3)$ that leads to the dynamical space-time structure and the generation of the electron mass.

The $\eta$-spinor components are the coefficient functions in the following decomposition
\begin{eqnarray}
D^i = \eta^0(\pi^1, \pi^2, \pi^3) P^i + \eta^1(\pi^1, \pi^2, \pi^3) J^i,
\end{eqnarray}
where $P^i$ is the energy-momentum vector and $J^i$ is the normal (in the sense of the Fubini-Study metric) Jacobi vector field
\cite{Besse}. Therefore
\begin{eqnarray}\label{}
\eta^0=\frac{G_{ik*} D^i P^{k*}}{G_{ik*} P^i P^{k*}},\quad
\eta^1=\frac{G_{ik*} D^i J^{k*}}{G_{ik*} J^i J^{k*}}.
\end{eqnarray}
Then under the comparison of LDV $D^i$, whose shift is
induced by the interaction used for a measurement, at two infinitesimally close GCS's (generalized coherent states) $(\pi^1,\pi^2,\pi^3)$ and
$(\pi^1+\delta^1,\pi^2+\delta^2,\pi^3+\delta^3)$ connected by a geodesic of $CP(3)$, one can get a nearby
spinor $(\eta^0+\delta \eta^0, \eta^1+\delta \eta^1)$ that may be calculated. Since $CP(3)$ is a totally geodesic manifold \cite{KN}, each geodesic belongs to some $CP(1)$ parameterized by the single complex coordinate $\pi=e^{-i\phi} \tan(\theta/2)$ will be used thereafter.
Thereby the Higgs potential will be replaced by the affine gauge potential (28)
\begin{figure}[h]
\begin{center}
    \includegraphics[width=4in]{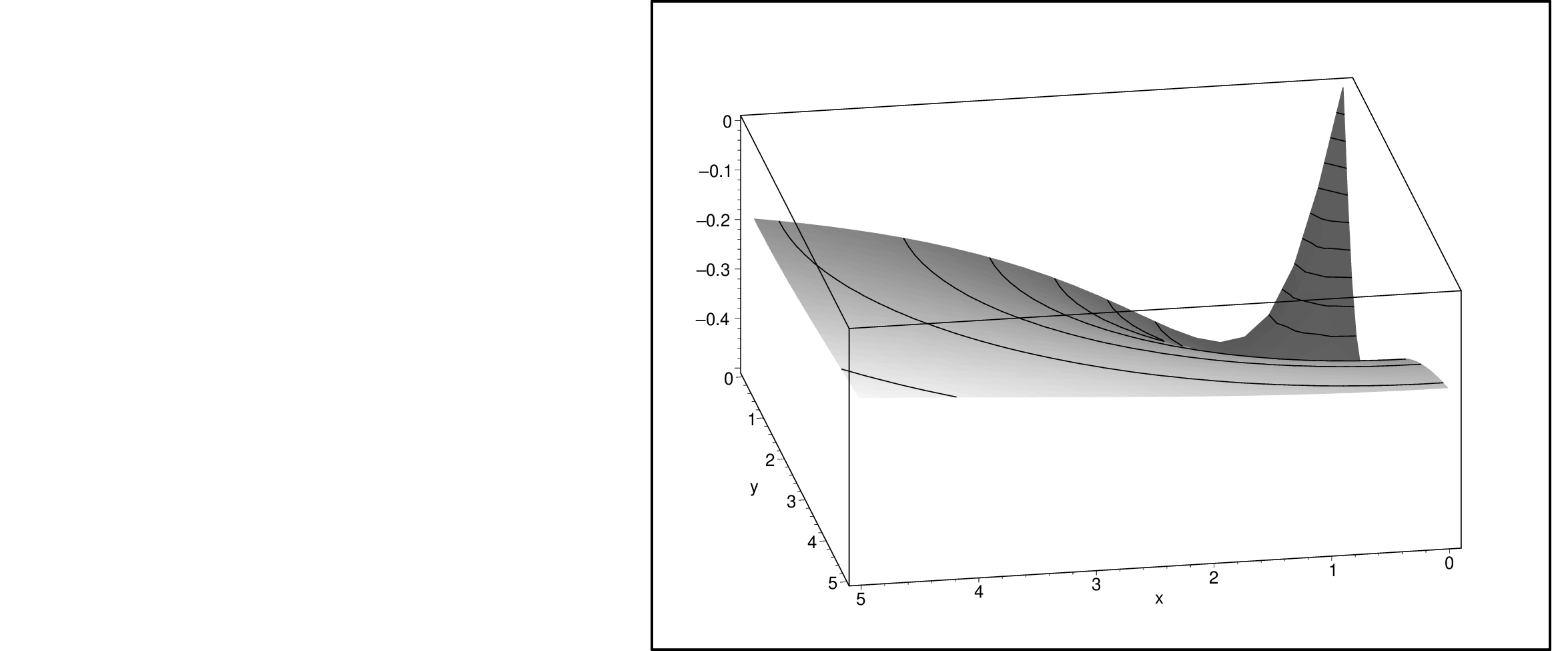}\\
  \caption{The shape of the gauge potential associated with the affine connection in CP(1): $\Gamma=-2\frac{|\pi|}{1+|\pi|^2},
\pi=x+iy. $}\label{fig.1}
  \end{center}
  \end{figure}
involved in the parallel transport of $D^i$ \cite{Le2,Le7} which agrees with the Fubini-Study metric (21).
It is worth while to note that this construction should solve ``the quantum measurement problem" in a natural and objective manner since the outcome of ``measurement" of LDV is provided by a spontaneous ``falling down" of the solution into the point $(\pi^1=\pi^2=\pi^3=0)$ or into the valley of the affine gauge potential. It means that under interaction used for the measurement of LDV $D^i$ spontaneously occurs (after elapse of the metabolic time) one of the qubit spinor component $\eta^0$ or $\eta^1$.

Two infinitesimally close qubit spinors $\eta$ and $\eta+\delta
\eta$ belonging to $C^2$ may be formally connected with infinitesimal $SL(2,C)$ transformations represented by ``Lorentz spin transformations
matrix'' \cite{G}
\begin{eqnarray}\label{31}
\hat{L}=\left( \begin {array}{cc} 1-\frac{i}{2}\delta \tau ( \omega_3+ia_3 )
&-\frac{i}{2}\delta \tau ( \omega_1+ia_1 -i ( \omega_2+ia_2)) \cr
-\frac{i}{2}\delta \tau
 ( \omega_1+ia_1+i ( \omega_2+ia_2))
 &1-\frac{i}{2}\delta \tau( -\omega_3-ia_3)
\end {array} \right).
\end{eqnarray}
I have assumed that there is not only formal but dynamical reasons, namely:
self-interaction of charge with the energy-momentum encoded by the dynamics
of two-level system whose components comprise the qubit spinor.
Therefore this process may be represented in the DST (dynamical space-time) associated with the manifold of coordinates in the attached Lorentz reference frame.
Then ``quantum accelerations" $a_1,a_2,a_3$ and ``quantum angular velocities" $\omega_1,
\omega_2, \omega_3$ may be found in the linear approximation from
the equation $\delta \eta = \hat{L} \eta-\eta$, or, strictly speaking, from
its consequence - the equations for the velocities $\xi^0=\frac{\delta \eta^0}{\delta \tau}$ and
$\xi^1=\frac{\delta \eta^1}{\delta \tau}$ of $\eta$ spinor variations
\begin{eqnarray}
\hat{R}\left(
  \begin{array}{cc}
    \eta^0  \cr
    \eta^1
  \end{array}
\right) =
\frac{1}{\delta \tau}(\hat{L}-\hat{1})\left(
  \begin{array}{cc}
    \eta^0  \cr
    \eta^1
  \end{array}
\right) = \left(
  \begin{array}{cc}
    \xi^0 \cr
    \xi^1
  \end{array}
\right).
\end{eqnarray}

If one puts $\pi=e^{-i\phi} \tan(\theta/2)$ then $\frac{\delta \pi}{\delta \tau}=
\frac{\partial \pi}{\partial \theta}\frac{\delta \theta}{\delta \tau}+
\frac{\partial \pi}{\partial \phi}\frac{\delta \phi}{\delta \tau}$, where
\begin{eqnarray}
\frac{\delta \theta}{\delta \tau}=-\omega_3\sin(\theta)-((a_2+\omega_1)\cos(\phi)+
(a_1-\omega_2)\sin(\phi))\sin(\theta/2)^2 \cr
-((a_2-\omega_1)\cos(\phi)+
(a_1+\omega_2)\sin(\phi))\cos(\theta/2)^2; \cr
\frac{\delta \phi}{\delta \tau}=a_3+(1/2)(((a_1-\omega_2)\cos(\phi)-
(a_2+\omega_1)\sin(\phi))\tan(\theta/2) \cr
-((a_1+\omega_2)\cos(\phi)-
(a_2-\omega_1)\sin(\phi))\cot(\theta/2)),
\end{eqnarray}
then one has the linear system of 6 real non-homogeneous equation
\begin{eqnarray}
\Re(\hat{R}_{00}\eta^0+\hat{R}_{01}\eta^1)&=&\Re(\frac{\delta \eta^0}{\delta \tau}), \cr
\Im(\hat{R}_{00}\eta^0+\hat{R}_{01}\eta^1)&=&\Im(\frac{\delta \eta^0}{\delta \tau}), \cr
\Re(\hat{R}_{10}\eta^0+\hat{R}_{11}\eta^1)&=&\Re(\frac{\delta \eta^1}{\delta \tau}),
\cr
\Im(\hat{R}_{10}\eta^0+\hat{R}_{11}\eta^1)&=&\Im(\frac{\delta \eta^1}{\delta \tau}),
\cr
\frac{\delta \theta}{\delta \tau}&=&F_1, \cr
\quad \frac{\delta \phi}{\delta \tau}&=&F_2,
\end{eqnarray}
giving $\vec{a}_Q(\eta^0,\eta^1,\theta, \phi,F_1,F_2),\vec{\omega}_Q(\eta^0,\eta^1,\theta, \phi,F_1,F_2)$ as the functions of ``measured" components of LDV $(\eta^0,\eta^1)$, the local coordinates of GCS $(\theta, \phi)$ or complex $\pi$, and 2 real
perturbation frequencies $(F_1, F_2)$ acting along and transversal to a geodesic in $CP(3)$.

The infinitesimal transition from one GCS of the electron to another is now
accompanied by dynamical transition from one Lorentz frame to another.
Thereby, infinitesimal Lorentz transformations define infinitesimal
``dynamical space-time'' coordinate variations. It is convenient to take
Lorentz transformations in the following form
\begin{eqnarray}
ct'&=&ct+(\vec{x} \vec{a}_Q) \delta \tau \cr
\vec{x'}&=&\vec{x}+ct\vec{a}_Q \delta \tau
+(\vec{\omega}_Q \times \vec{x}) \delta \tau
\end{eqnarray}
where I put
$\vec{a}_Q=(a_1/c,a_2/c,a_3/c), \quad
\vec{\omega}_Q=(\omega_1,\omega_2,\omega_3)$ \cite{G} in order to have
for $\tau$ the physical dimension of time. The expression for the
``4-velocity" $ V^{\mu}$ is as follows
\begin{equation}\label{}
V^{\mu}_Q=\frac{\delta x^{\mu}}{\delta \tau} = (\vec{x} \vec{a}_Q,
ct\vec{a}_Q  +\vec{\omega}_Q \times \vec{x}) .
\end{equation}
The coordinates $x^\mu$ of the imaging point in dynamical space-time serve here merely for the parametrization of the energy-momentum distribution in the ``field
shell'' arising under ``morphogenesis" described by quasi-linear field
equations \cite{LeHor,Le3} in DST.

\section{Derivation of the ``field-shell" equations for non-local quantum electron}
The ``field-shell" equations may be derived as the consequence of the
conservation law of the energy-momentum \cite{Le2,Le3,LeHor}. In the reply on
questions of some colleagues (why, say, Lagrangian is not used for derivation of
the field equations?) I would like to note following. Strictly speaking the least
action principle is realized only in average that is clear
from Feynman's summation of quantum amplitudes. Hence one may suspect that
more deep principle should be used for derivation of fundamental equations of motion. The quantum formulation of the inertia law has been used \cite{Le1}.

What the inertial principle means for quantum systems and their states?
Formally the inertial
principle is tacitly accepted in the package with relativistic invariance.
But we already saw that the
problem of identification and therefore the localization of quantum particles
in classical space-time is problematic and it requires a clarification.

Quantum lump of non-local electron should presumably serve as extended source
of electromagnetic field. The dynamics of spin/charge degrees of freedom
may be mapped onto dynamical space-time
if one assumes that transition from one GCS of the electron to another is
accompanied by dynamical transition from one Lorentz frame to another.
This reflects the reconstruction of broken $SU(4)$ symmetry to $SL(2,C)$ of DST.
Thereby, infinitesimal Lorentz transformations creates small
``dynamical space-time'' coordinates variations parameterizing
energy-momentum distribution.

Since the operator of energy-momentum represented by a tangent vector field
to $CP(3)$ replaces the Lagrangian, I use a complex covariant differentiation
relative Fubini-Study metric instead of variation. Differential space-time
field equations arose in a section of tangent fibre bundle over $CP(3)$.
It leads naturally to some ``lump" solutions which should be carefully studied. I show here some preliminary results promising progress in understanding structure of the quantum electron. In particular it is clear that the quantum nature of derived field quasi-linear PDE's  (without references to classical analogy) would shed light on the their generic connection with Hamilton-Jacobi classical equations and de Broglie-Schr\"odinger optics-mechanics analogy.

Lets discuss the formulation of the quantum inertia law in the case of non-local
quantum electron \cite{Le1}. I assume that quantum version of the inertia law may be formulated as follows:

\textbf{inertial quantum motion of quantum system may be expressed as a
 self-conservation of its local dynamical variables like energy-momentum, spin, charge, etc.}

The conservation law of the energy-momentum vector field in $CP(3)$ during
inertial evolution will be expressed by the equation of the affine parallel transport
\begin{equation}
\frac{\delta [P^{\mu}\Phi_{\mu}^i(\gamma_{\mu})]}{\delta \tau}=0,
\end{equation}\label{32}
which is equivalent to the following system of four coupled quasi-linear PDE
for dynamical space-time distribution of energy-momentum ``field-shell" of
quantum state
\begin{equation}
V^{\mu}_Q (\frac{\partial
P^{\nu}}{\partial x^{\mu} } + \Gamma^{\nu}_{\mu \lambda}P^{\lambda})=
-\frac{c}{\hbar}(\Gamma^m_{mn} \Phi_{\mu}^n(\gamma) + \frac{\partial
\Phi_{\mu}^n (\gamma)}{\partial \pi^n}) P^{\nu}P^{\mu},
\end{equation}\label{33}
and ordinary differential equations for relative amplitudes
\begin{equation}
\quad \frac{d\pi^k}{d\tau}= \frac{c}{\hbar}\Phi_{\mu}^k P^{\mu},
\end{equation}\label{34}
which is in fact the \emph{equations of characteristic} for linear ``super-Dirac"
equation
\begin{equation}\label{35}
i P^{\mu}\Phi_{\mu}^i(\gamma_{\mu})\frac{\partial \Psi}{\partial \pi^i} =mc \Psi
\end{equation}
that supposes ODE for single ``total state function"
\begin{equation}\label{36}
i \hbar \frac{d \Psi}{d \tau} =mc^2 \Psi
\end{equation}
with the solution for variable mass $m(\tau)$
\begin{equation}\label{37}
\Psi(T) = \Psi(0)e^{-i\gamma_C} e^{-i\frac{c^2 }{\hbar}\int_0^T m(\tau) d\tau}.
\end{equation}

The system of quasi-linear PDE's following from the conservation law, ODE's
and algebraic linear non-homogeneous equations
comprise of the self-consistent problem for stability (in fact - existing) of LDV.

The solution of the ``field-shell" equations have been discussed
\cite{Le2}. The theory of these equations is well known \cite{Courant}.
One has the quasi-linear PDE system with identical principle part $V^{\mu}_Q$
for which we will build characteristics for the system of implicit solutions
for 4+4 extended variables
\begin{eqnarray}
\phi^1(x^0,x^1,x^2,x^3,P^0,P^1,P^2,P^3)=c^1;\cr
\phi^2(x^0,x^1,x^2,x^3,P^0,P^1,P^2,P^3)=c^2;\cr
\phi^3(x^0,x^1,x^2,x^3,P^0,P^1,P^2,P^3)=c^3;\cr
\phi^4(x^0,x^1,x^2,x^3,P^0,P^1,P^2,P^3)=c^4.
\end{eqnarray}\label{39}
Differentiation of $\phi^{\mu}$ in $x^{\nu}$ gives
\begin{equation}\label{40}
\frac{\partial
\phi^{\mu}}{\partial x^{\nu} }  + \frac{\partial
\phi^{\mu}}{\partial P^{\lambda}}(\frac{\partial
P^{\lambda}}{\partial x^{\nu}}+ \Gamma^{\lambda}_{\nu \mu}P^{\mu})=0.
\end{equation}
This equation being multiplied by
$\frac{\delta x^{\nu}}{\delta \tau} = V^{\nu}_Q$ gives the equation
\begin{equation}
\frac{\delta
\phi^{\mu}}{\delta \tau }=\frac{\partial
\phi^{\mu}}{\partial x^{\nu} }\frac{\delta x^{\nu}}{\delta \tau}  + \frac{\partial
\phi^{\mu}}{\partial P^{\lambda}}(\frac{\partial
P^{\lambda}}{\partial x^{\nu}}+ \Gamma^{\lambda}_{\nu \mu}P^{\mu})
\frac{\delta x^{\nu}}{\delta \tau}=0
\end{equation}\label{41}
or
\begin{eqnarray}
\frac{\partial
\phi^{\mu}}{\partial x^{\nu} } V^{\nu}_Q + \frac{\partial
\phi^{\mu}}{\partial P^{\lambda}}(\frac{\partial
P^{\lambda}}{\partial x^{\nu}}+ \Gamma^{\lambda}_{\nu \mu}P^{\mu})
V^{\nu}_Q \cr  = \frac{\partial
\phi^{\mu}}{\partial x^{\nu} } V^{\nu}_Q - \frac{\partial
\phi^{\mu}}{\partial P^{\lambda}}\frac{c}{\hbar}(\Gamma^m_{mn}
\Phi_{\mu}^n(\gamma) + \frac{\partial
\Phi_{\mu}^n (\gamma)}{\partial \pi^n}) P^{\lambda}P^{\mu}=0.
\end{eqnarray}\label{42}
Redefinition of the coefficients $C^{\nu +\lambda}:=-\frac{c}{\hbar}
(\Gamma^m_{mn} \Phi_{\mu}^n(\gamma) + \frac{\partial
\Phi_{\mu}^n (\gamma)}{\partial \pi^n}) P^{\lambda}P^{\mu}$ and variables
$x^{\nu +\lambda}:=P^{\lambda}$ gives a possibility to rewrite this
equation for any  $\phi = \phi_{\mu}$ as follows
\begin{equation}
\sum_{\kappa=1}^8 C^{\kappa} \frac{\partial
\phi}{\partial x^{\kappa}}=0.
\end{equation}\label{43}
Then one has the system of ODE's of characteristics
\begin{eqnarray}
\frac{\delta x^{\nu}}{\delta \tau}&=&V^{\nu}_Q,\cr
\frac{\delta P^{\nu}}{\delta \tau}&=&-V^{\mu}_Q
\Gamma^{\nu}_{\mu \lambda}P^{\lambda}-\frac{c}{\hbar}(\Gamma^m_{mn}
\Phi_{\mu}^n(\gamma) + \frac{\partial
\Phi_{\mu}^n (\gamma)}{\partial \pi^n}) P^{\nu}P^{\mu}, \cr
\frac{d\pi^k}{d\tau} &=& \frac{c}{\hbar}\Phi_{\mu}^k P^{\mu}.
\end{eqnarray}\label{44}

\section{Solutions of ``field-shell" equations for self-interacting electron}
I will discuss now the solution of the ``field-shell" equations (42).
The integration of a ``cross" combination of the characteristic equations from the first and the second system (52) should be done. One of the combination is as follows
\begin{eqnarray}
\frac{\delta x^0}{V^0_Q}=\frac{\delta P^0}{P^0(L_0P^0+L_1P^1+L_2P^2+L_3P^3)},
\end{eqnarray}
where $L_{\mu}=-\frac{c}{\hbar}(\Gamma^m_{mn} \Phi_{\mu}^n(\gamma) + \frac{\partial
\Phi_{\mu}^n (\gamma)}{\partial \pi^n})$. If $L_0P^0<0$ then one has implicit solution
\begin{eqnarray}
\frac{x^0}{a_{\alpha}x^{\alpha}} + T^0=-\frac{2}{L_{\alpha}P^{\alpha}}\tanh^{-1}(1+\frac{2L_0P^0}{L_{\alpha}P^{\alpha}}),
\end{eqnarray}
where $T^0$ is an integration constant. An explicit solution for energy is the kink
\begin{eqnarray}
P^0=\frac{L_{\alpha}P^{\alpha}}{2L_0}[ \tanh(-(\frac{x^0}{a_{\alpha}x^{\alpha}} + T^0)\frac{L_{\alpha}P^{\alpha}}{2})-1].
\end{eqnarray}
If I put $\frac{L_{\alpha}P^{\alpha}}{2}=1, L_0=1, V=a_{\alpha}x^{\alpha}=0.6$  the kink solution may be represented by the graphic in Fig. 2.
\begin{figure}[h]
  \begin{center}
    \includegraphics[width=4in]{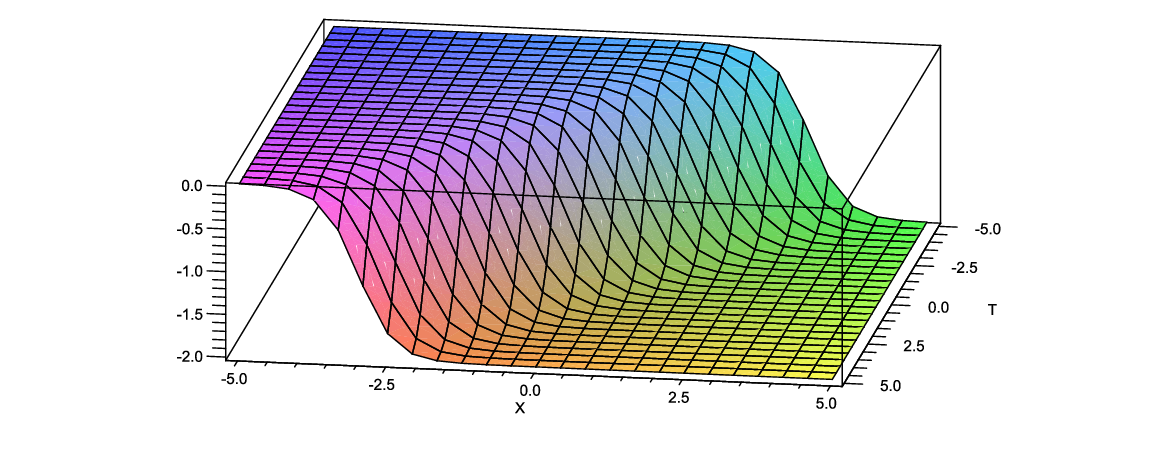}\\
  \caption{The kink solution (55) of (53). }\label{fig.2}
  \end{center}
  \end{figure}

This solution represent the lump of an electron self-interacting through an electro-magnetic-like field in the co-moving Lorentz reference frame. The nature of this field will be discussed in a separate article. The lump ``modulates" the ordinary plane wave essentially only in the vicinity of the core of the lump. In  standard QED self-interacting effects are treated as a polarization of the vacuum. In the present picture the lump is dynamically self-supporting by outward and inward waves whose characteristics are represented by the equations (49).

\section{Stability of energy-momentum characteristics and dispersion law}
Let me discuss the stability of energy-momentum characteristics given by the system of four ODE's
\begin{eqnarray}
\frac{\delta P^{\lambda}}{\delta \tau}&=&-V^{\mu}_Q  \Gamma^{\nu}_{\mu \lambda}P^{\lambda}-\frac{c}{\hbar}(\Gamma^m_{mn} \Phi_{\mu}^n(\gamma) + \frac{\partial
\Phi_{\mu}^n (\gamma)}{\partial \pi^n}) P^{\nu}P^{\mu}.
\end{eqnarray}
If we seriously treat these characteristics as trajectories of electrons in $CP(3)$ then their stability in DST is an essential problem. The self-interaction electron is represented here as a dynamical field system whose equilibrium is provided by the counterbalance of outward and inward waves.

The standard approach to stability analysis instructs us to find the stationary points. The stationary condition
\begin{eqnarray}
\frac{\delta P^{\lambda}}{\delta \tau}&=&0
\end{eqnarray}
leads to the system of algebraic equations
\begin{eqnarray}
V^{\mu}_Q  \Gamma^{\nu}_{\mu \lambda}P^{\lambda}+\frac{c}{\hbar}(\Gamma^m_{mn} \Phi_{\mu}^n(\gamma) + \frac{\partial
\Phi_{\mu}^n (\gamma)}{\partial \pi^n}) P^{\nu}P^{\mu}=0.
\end{eqnarray}
Let me investigate initially the simplified case neglecting the space-time connection term  $V^{\mu}_Q  \Gamma^{\nu}_{\mu \lambda}P^{\lambda}$. This gives us more simple equations for stationary points
\begin{eqnarray}
(\Gamma^m_{mn} \Phi_{\mu}^n(\gamma) + \frac{\partial
\Phi_{\mu}^n (\gamma)}{\partial \pi^n}) P^{\nu}P^{\mu}=0.
\end{eqnarray}
telling us that in the non-trivial case (i.e. exclusion of condition $P^{\nu}=0$), one has the equation of hyper-plane
\begin{eqnarray}
(\Gamma^m_{mn} \Phi_{\mu}^n(\gamma) + \frac{\partial
\Phi_{\mu}^n (\gamma)}{\partial \pi^n}) P^{\mu}_0=0
\end{eqnarray}
``rotating" with variation of local coordinates $\pi^i$. The probing solution in the vicinity of the stationary points $P^{\mu}_0$ is as follows
\begin{eqnarray}
P^{\mu}(\tau)=P^{\mu}_0 + p^{\mu} e^{\omega \tau}.
\end{eqnarray}
This solution being substituted in the equation
\begin{eqnarray}
\frac{\delta P^{\nu}}{\delta \tau}&=&-\frac{c}{\hbar}(\Gamma^m_{mn} \Phi_{\mu}^n(\gamma) + \frac{\partial
\Phi_{\mu}^n (\gamma)}{\partial \pi^n}) P^{\nu}P^{\mu}.
\end{eqnarray}
leads to the linear system
\begin{eqnarray}
\frac{\hbar \omega}{c}p^{\nu}+ (\Gamma^m_{mn} \Phi_{\mu}^n(\gamma) + \frac{\partial
\Phi_{\mu}^n (\gamma)}{\partial \pi^n}) p^{\mu}P^{\nu}_0 = 0.
\end{eqnarray}
The determinant of this homogeneous liner system is as follows
\begin{eqnarray}
D= (\frac{\hbar \omega}{c})^4+(\frac{\hbar \omega}{c})^3(\Gamma^m_{mn} \Phi_{\mu}^n(\gamma) + \frac{\partial
\Phi_{\mu}^n (\gamma)}{\partial \pi^n}) P^{\mu}_0.
\end{eqnarray}
The determinant put to zero together with the
hyper-plane equation gives $\omega=0$. This means that the mass of self-interacting quantum electron is zero for trivial space-time flat connection.

I would like to compare the ``off-shell" dispersion law (55)
with the de Broglie ``on-shell" dispersion law. The result has been shown in Fig. 3.
\begin{figure}[h]
  \begin{center}
    \includegraphics[width=4in]{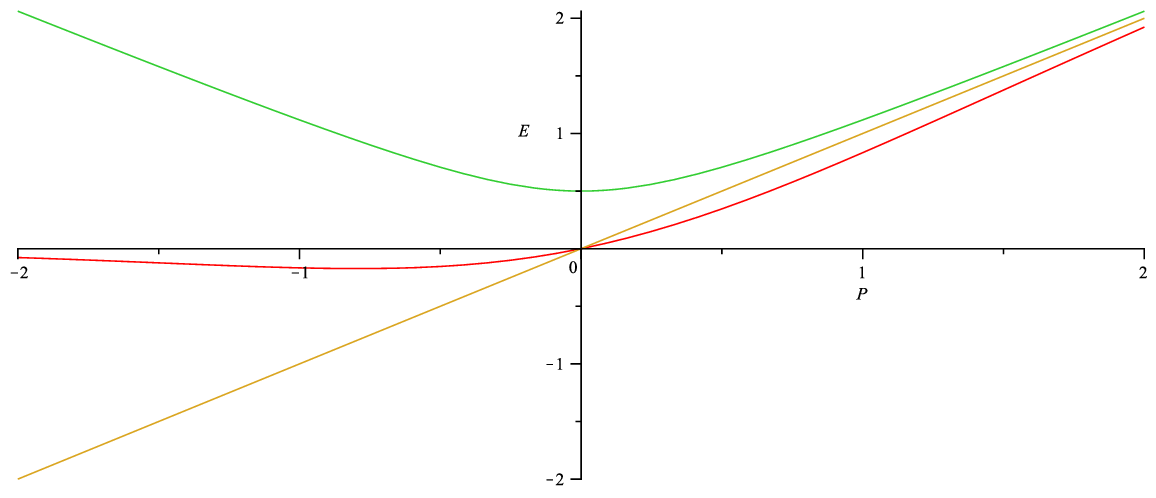}\\
  \caption{``Off-shell" dispersion law (red) in comparison with de Broglie ``on-shell" dispersion law (green) and asymptote $E=cP$ (brown).}\label{fig.3}
  \end{center}
  \end{figure}
It is clearly seen that it traverses below the asymptote $E=cP$, whereas de Broglie ``on-shell" dispersion law $P^{\mu}P_{\mu}-m^2c^2=0$ traverses above it.
The comparison of our dispersion law with Blochintzev spectrum
 $E^2=P^2-m^2c^2$ is shown Fig. 4.
\begin{figure}[h]
  \begin{center}
    \includegraphics[width=4in]{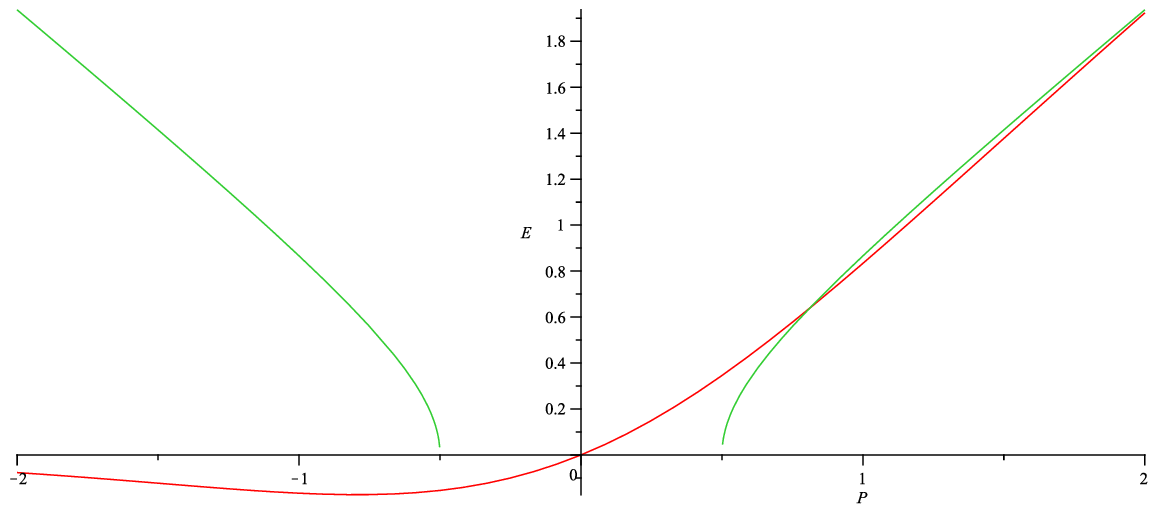}\\
  \caption{``Off-shell" dispersion law (red) in comparison with Blochintzev
   dispersion law of two strongly coupled linear oscillators (green).}\label{fig.4}
  \end{center}
  \end{figure}
The approximate expression for dispersion in the vicinity of zero is as follows
\begin{eqnarray}
E_{app}=\frac{c}{2}P+\frac{c^2}{4\hbar}(T_0+\frac{x_0}{V})P^2 \cr =\frac{c^2}{4\hbar}(T_0+\frac{x_0}{V})[(P+\frac{\hbar}{c}\frac{V}{x_0+VT_0})^2
-(\frac{\hbar}{c}\frac{V}{x_0+VT_0})^2]
\end{eqnarray}
and it is depicted in Fig. 5.
\begin{figure}[h]
  \begin{center}
    \includegraphics[width=4in]{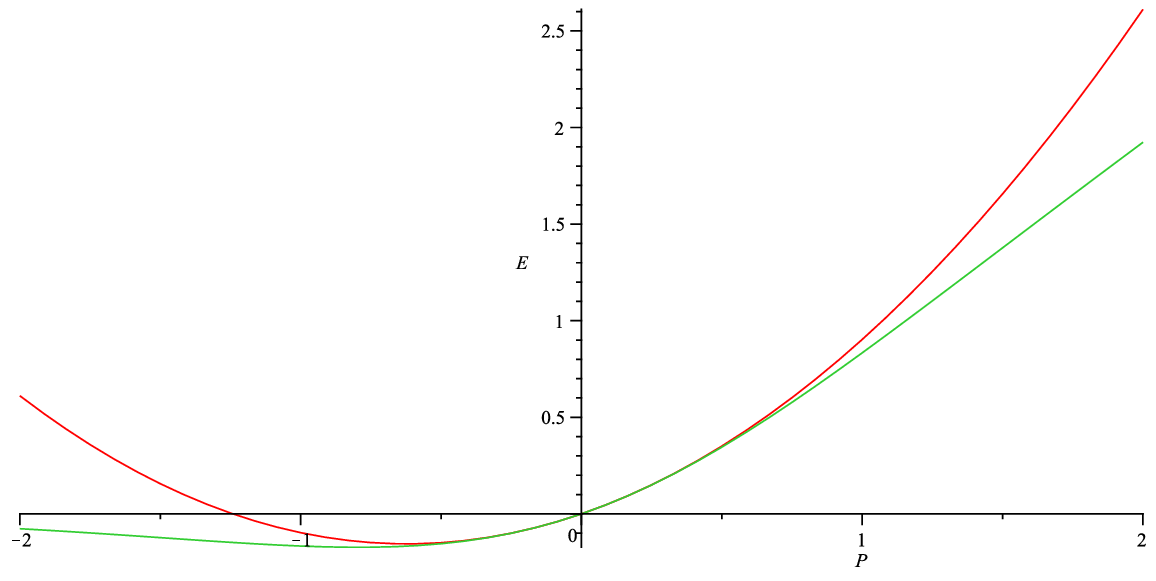}\\
  \caption{Approximate square dispersion law (red) in comparison with ``off-shell"
  dispersion law (green) in the vicinity of zero.}\label{fig.5}
  \end{center}
  \end{figure}
The minimum of the energy is as follows
\begin{eqnarray}
E_{min}=-\frac{\hbar}{4}\frac{V}{x_0+VT_0}
\end{eqnarray}
at the momentum
\begin{eqnarray}
P_{min}=-\frac{\hbar}{c}\frac{V}{x_0+VT_0}.
\end{eqnarray}
The group velocity of propagation shows that there is the a ``zone" of wave vectors where one has a space-like leakage of self-interacting field, see Fig. 6.
\begin{figure}[h]
  \begin{center}
    \includegraphics[width=4in]{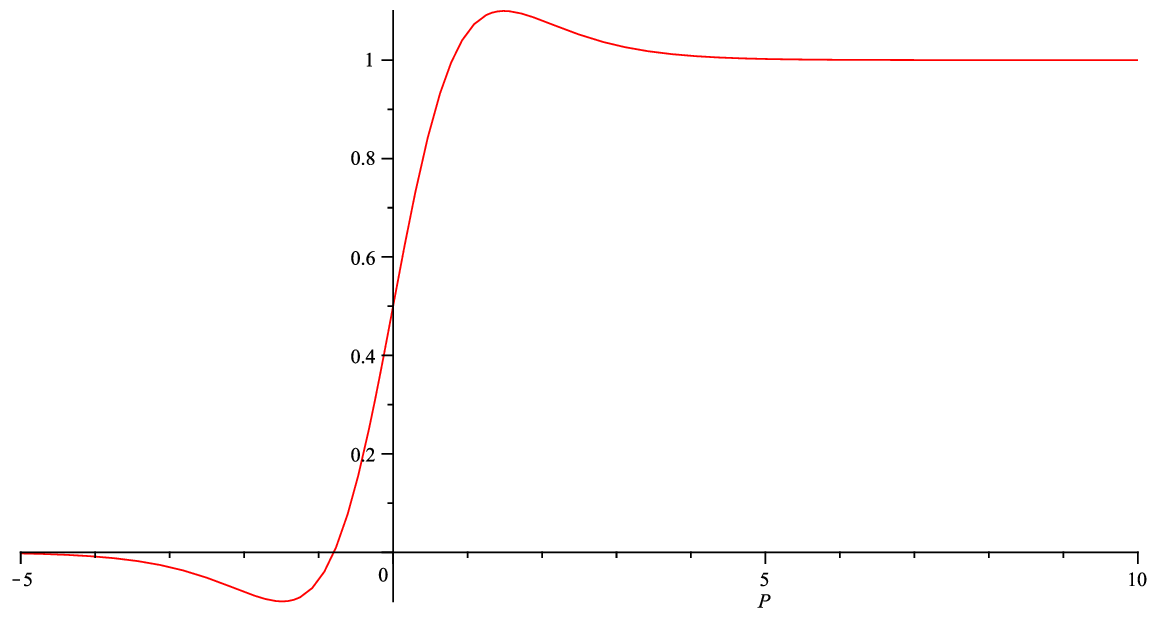}\\
  \caption{The group velocity of self-interacting field.}\label{fig.6}
  \end{center}
  \end{figure}
It may be related to the quantum entanglement but this phenomenon may be naturally realized in the context of multi-kink solutions and it will be studied later. Only asymptotically it tends to the velocity of light together with phase velocity, see Fig. 7. It really appears to be like the ``unparticle" sector of Blochintzev:
phase velocity is always smaller than $c$, but the behavior of group velocity is
more complicated.
\begin{figure}[h]
  \begin{center}
    \includegraphics[width=4in]{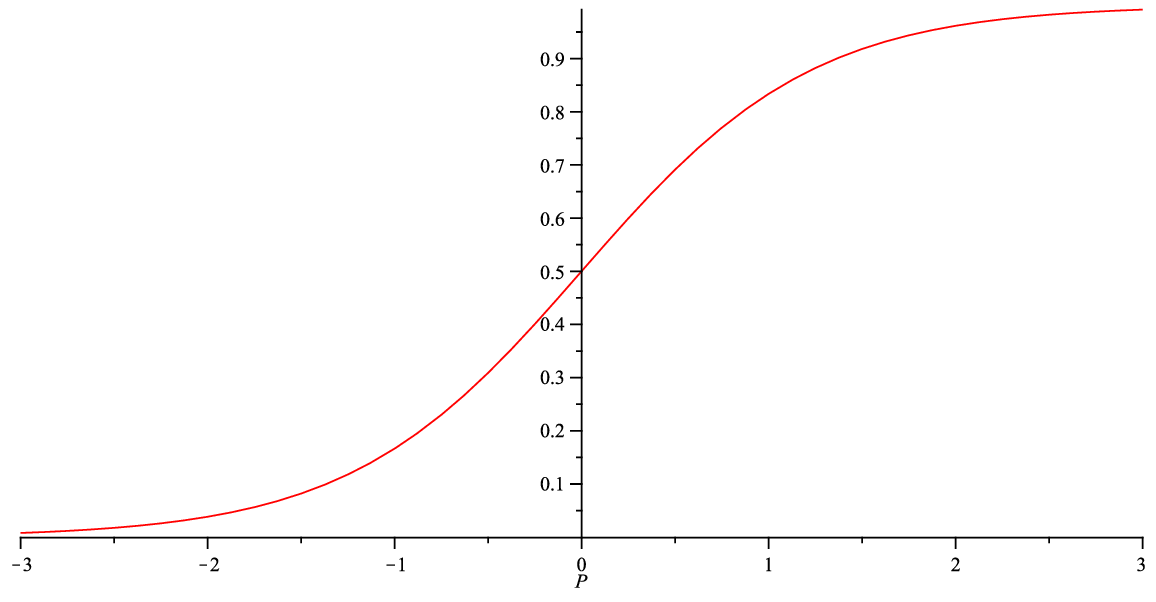}\\
  \caption{The phase velocity of self-interacting field.}\label{fig.7}
  \end{center}
  \end{figure}

I would like to note that  $V=x^{\alpha}a_{\alpha}=\vec{x} \vec{a}_Q$ is the time component of the ``4-velocity" (39)
proportional to the ``quantum acceleration" $\vec{a}_Q$. This boost parameter together with angular velocity $\vec{\omega}_Q$ is the solution of the non-homogeneous system of linear equations (38) expressing the condition of measurability of LDV. In fact it is the condition of existence for the numerical value of LDV expressed by complex vector field in $CP(3)$. The boost and angular rotation parameters depend on the affine connection associated with gauge potential in $CP(3)$. It gives some hope for treating the expression  $(P+\frac{\hbar}{c}\frac{V}{x_0+VT_0})^2$ in $E_{app}$ as the momentum with an additional ``electromagnetic-like" potential generated by the geometry of $SU(4)$ sub-manifold $CP(3)$.
The complex scalar potential of electromagnetic field generated by the logarithm of the dimensionless projective coordinate has been already discussed \cite{Friedman}. The authors treated the electric field as a generator of a boost and the magnetic field as the generator of rotations. The quantum conditions of measurability of LDV (38) invert in fact this approach: the geometry of the coset manifold $CP(3)$ and generators of $SU(4)$ expressed in local coordinates $(\pi^1, \pi^2, \pi^3)$
create ``electromagnetic-like" field. This question is not properly understood yet and it requires an additional investigation concerning the sectional curvature of the $CP(3)$.

The gapless dispersion law discussed above arose in the flat Minkowski space-time. It is a consequence of simplified condition of characteristic stability when the space-time connection is trivial. Therefore in order to find the ``optical" dispersion law with a mass-gap and state-dependent attractor corresponding to finite mass of the electron one should analyse the full equation (56). Then we come to the homogeneous linear system
\begin{eqnarray}
\frac{\hbar \omega}{c}p^{\nu}+\frac{\hbar }{c}V^{\mu}_Q  \Gamma^{\nu}_{\mu \lambda}p^{\lambda}+ (\Gamma^m_{mn} \Phi_{\mu}^n(\gamma) + \frac{\partial
\Phi_{\mu}^n (\gamma)}{\partial \pi^n}) p^{\mu}P^{\nu}_0 = 0.
\end{eqnarray}
The determinant of this system is as follows
\begin{eqnarray}
D_1= (\frac{\hbar \omega}{c})^4+\alpha (\frac{\hbar \omega}{c})^3+\beta (\frac{\hbar \omega}{c})^2+\gamma (\frac{\hbar \omega}{c})+\delta,
\end{eqnarray}
with complicated coefficients $\alpha, \beta, \gamma, \delta$. I put $K^{\nu}_{\lambda}=\frac{\hbar }{c}V^{\mu}_Q  \Gamma^{\nu}_{\mu \lambda}$ and $M^{\nu}_{\mu}=(\Gamma^m_{mn} \Phi_{\mu}^n(\gamma) + \frac{\partial
\Phi_{\mu}^n (\gamma)}{\partial \pi^n}) P^{\nu}_0$ then
one may find that
\begin{eqnarray}
\alpha=Tr(K^{\nu}_{\lambda})+ Tr(M^{\nu}_{\mu})
\end{eqnarray}
and
\begin{eqnarray}
\beta=K^0_0(L_1P_0^1+L_2P_0^2+L_3P_0^3)+K^1_1(L_0P_0^0+L_2P_0^2+L_3P_0^3) \cr +K^2_2(L_1P_0^1+L_0P_0^0+L_3P_0^3)+K^3_3 (L_1P_0^1+L_0P_0^0+L_2P_0^2)\cr
-K^0_1 L_0P_0^1-K^1_0 L_1P_0^0 -K^0_2 L_0P_0^2-K^2_0 L_2P_0^0
-K^0_3 L_0P_0^3-K^3_0 L_3P_0^0 \cr -K^1_2 L_1P_0^2-K^2_1 L_2P_0^1
-K^1_3 L_1P_0^3-K^3_1 L_3P_0^1 -K^2_3 L_2P_0^3-K^3_2 L_3P_0^2,
\end{eqnarray}
whereas $ \gamma, \delta$ have higher order in $G_N$ and they may be temporarily  discarded in our approximate dispersion law.
This dispersion law may be written as follows
\begin{eqnarray}
 (\frac{\hbar \omega}{c})^2[(\frac{\hbar \omega}{c})^2+\alpha (\frac{\hbar \omega}{c})+\beta] =0.
\end{eqnarray}
The trivial solution $\omega_{1,2}=0$ has already been discussed. Two non-trivial solutions in weak gravitation field when $\alpha^2 \gg \beta$ are given by the equations
\begin{eqnarray}
 \hbar \omega_{3,4} =c \alpha \frac{-1 \pm \sqrt{1-\frac{4\beta}{\alpha^2}}}{2} \approx c \alpha \frac{-1 \pm (1-\frac{2\beta}{\alpha^2})}{2};\cr
 \hbar \omega_3=\frac{-c \beta}{\alpha}, \quad \hbar \omega_4=-c \alpha+\frac{c \beta}{\alpha}.
 \end{eqnarray}
Both parameters $\alpha, \beta$ are in fact complex functions of $(\pi^1,\pi^2,\pi^3)$ but they are linear in stationary momenta $P^{\mu}_0$.
Coefficients $K^{\nu}_{\mu} \propto G_N$ and generally they are much smaller than $M^{\nu}_{\mu}$. The  negative real part of these two roots substituted in the probing function (65) will define attractors and two finite masses.

Lets find initially solution of the non-linear system (61).
Its approximate solution in the vicinity of $P^{\mu}_{test} = (mc^2, 0, 0, 0) $
has been found by the method of Newton:
\begin{eqnarray}
P_0^{\mu} =P^{\mu}_{test} + \delta^{\mu}+...,
\end{eqnarray}\label{55}
where $\delta^{\mu}$ is the solution of the Newton's first approximation equations
\begin{eqnarray}
(2L_0mc+K^0_0)\delta^0+(L_1mc+K^0_1)\delta^1+\cr
(L_2mc+K^0_2)\delta^2+(L_3mc+K^0_3)\delta^3 &=& -\frac{(L_0m^2c^4+K^0_0mc^3)}{c^2} \cr
K^1_0\delta^0+(L_0mc+K^1_1)\delta^1+K^1_2\delta^2+K^1_3\delta^3 &=& -K^1_0mc \cr
K^2_0\delta^0+K^2_1\delta^1+(L_0mc+K^2_2)\delta^2+K^2_3\delta^3 &=& -K^2_0mc \cr
K^3_0\delta^0+K^3_1\delta^1+K^3_2\delta^2+(L_0mc+K^3_3)\delta^3 &=& -K^3_0mc,
\end{eqnarray}\label{56}
where $L_{\mu}=(\Gamma^m_{mn} \Phi_{\mu}^n(\gamma) + \frac{\partial
\Phi_{\mu}^n (\gamma)}{\partial \pi^n})$ is now dimensionless.

It has been assumed that self-interaction of charge and spin degrees of
freedom comprise the energy-momentum whose distribution is encoded by field
dynamics in dynamical space-time (DST) with help of two-level system represented
by the qubit spinor \cite{Le2}. This DST will be associated with manifold
of coordinates in Lorentz reference frame attached to LDV during the virtual
``measurement".

If hypothesis about dynamical nature of electron mass defined by self-interacting
spin/charge degrees of freedom is correct then it is very natural to assume that
\begin{eqnarray}\label{61}
F_1 = \frac{\delta \theta}{\delta \tau} =\Re(\omega_{3}) =
\frac{c}{\hbar}\Re(\frac{- \beta}{ \alpha}),or \cr
F_1 = \frac{\delta \theta}{\delta \tau} =\Re(\omega_{4}) =
\frac{c}{\hbar} \Re(- \alpha+\frac{ \beta}{\alpha}), and \cr
F_2 = \frac{\delta \phi}{\delta \tau} =\Im(\omega_{3})
=\frac{c}{\hbar}\Im(\frac{- \beta}{ \alpha}),or \cr
F_2 = \frac{\delta \phi}{\delta \tau} =\Im(\omega_{4})
=\frac{c}{\hbar} \Im(- \alpha+\frac{ \beta}{\alpha}).
\end{eqnarray}
Solution of complicated self-consistent problem (38), (52), (75), (76)
is not found yet. This ``field-shell" solution
for the self-interacting quantum electron should contain dynamically generated mass.

\section{Conclusion}
Primarily, there were two mathematical approaches to the formulation of quantum theory. The first one (developed by Hiesenberg) makes accent on the non-commutative character of new ``quantized" dynamical variables whereas the second one (developed by Schr\"odinger) replaces ordinary differential Hamilton's equations of classical dynamics by linear differential equations in partial derivatives associated with Hamilton-Jacobi equation \cite{Sch1}. Both approaches are equivalent in the framework of so-called optics-mechanics analogy and comprise the basis for modern quantum mechanics. This analogy, however, is limited by itself for very clear reasons: mechanics is merely a coarse approximation (even being generalized to many-dimension dynamics of Hertz) and the ``optics" of the action waves is too tiny for description of the complicated structure of ``elementary" quantum particles. This was realized already during the first attempts to synthesize relativistic and quantum principles.

Analysis of the foundations of quantum theory and relativity shows that there are two types of symmetries. One of them is the symmetry of relative space-time transformations of the whole setup which reflects, say, the \emph{first order of relativity}. A different type of (state dependent) symmetry is realized in the quantum state space relative to local infinitesimal variation of a flexible setup (\emph{second order of relativity or ``super-relativity"} \cite{Le1,Le2,Le3}). Gauge invariance is a particular case of this type of symmetry. Analysis shows that it is impossible to use ordinary primordial elements like particles, material points, etc., trying to build a consistent theory. Even space-time cannot conserve its independent and a priori structure. Therefore the unification of relativity and quantum principles may be formalized if one uses new primordial elements and the classification of their motions: rays of quantum states instead of material points (particles) and complex projective Hilbert state space $CP(N-1)$ where these states move under the action of the unitary group $SU(N)$ instead of space-time \cite{Le2,Le3}.

Then:

1. Dynamical variables are in fact the generators of the group of symmetry and their non-commutative character is only a consequence of the curvature of the group manifold \cite{G}. State-dependent realization of $SU(N)$ generators as vector fields on $CP(N-1)$ evidently reveals the non-trivial global geometry of $SU(N)$ and its coset sub-manifold \cite{Le2,Le4}.

2. Attempts ``to return" in the Minkowski space-time (after second quantization) from the Schr\"odinger's configuration space is successful for statistical aims but they are not consistent on the fundamental level of a single quantum particle (which without any doubt does exist!) and therefore should be revised. In fact initially one should \emph{delete global space-time} by transition to a ``co-moving frame" and after virtual infinitesimal displacement of the generalized coherent state (GCS) of the electron \emph{to restore state-dependent local dynamical space-time}.

3.  The physically correct transition from quantum to classical mechanics arose as a serious problem immediately after the formulation of the wave mechanics of Schr\"odinger \cite{Sch2}. The failure to build stable wave packet for single electron from solutions of linear PDE's led to the probabilistic interpretation of the wave function. Further progress in the theory of non-linear PDE's like  sin-Gordon or KdV renewed generally the old belief in the possibility to return to non-singular quantum particles \cite{Rajaraman}.

The revision mentioned above (see point (2)) proposed here is intended to derive new non-linear quantum equations for a self-interacting nonlocal electron.
Notice that new field equations could not contain an arbitrary potential as it was in the case of Schr\"odinger or Dirac equations. This potential should be generated by the spin/charge self-interaction. One of the consistent prosedure is to use quasi-linear field PDE's following from the conservation laws that has been already discussed \cite{Le1,Le2,Le3,Le4,Le5,Le6}. It is provided by a state dependent local non-Abelian ``chiral" gauge field acting on $CP(3)$ as a tangent vector field.

Perturbation of a generalized coherent state of $G=SU(4)$ of the electron is studied in the vicinity of the stationary degenerated state given by ordinary (not secondly quantized) Dirac equation. This perturbation is generated by coset transformations $G/H=SU(4)/S[U(1) \times U(3)]=CP(3)$ as an analog of the infinitesimal F.-W. transformations \cite{Le1}. Self-interaction arises due to the curvature of the projective Hilbert space $CP(3)$ and the state-dependent dynamical space-time (DST) is built during ``objective quantum measurement" \cite{Le3}.
\section{Summary}
A new model of a non-local self-interacting quantum electron has been proposed.
Such self-interaction is provided by the spin-charge quantum dynamics. The non-linear realization of the $\gamma$-matrices of Dirac by the tangent vector fields to $CP(3)$ is used instead of the second quantization. The back-reaction of the internal dynamics reflects in ``slow" accelerated motion of the attached ``Lorentz reference frame" introducing state-dependent dynamical space-time coordinates. A ``field-shell" of energy-momentum distribution is described by the system of quasi-linear PDE's. These are the consequence of the conservation law of energy-momentum vector field expressed by affine parallel transport in $CP(3)$ which agrees with the Fubini-Study metric.

``Off-shell" dispersion law, group and phase velocities asymptotically coincide
with de Broglie ``on-shell" dispersion law. These excitations of the self-interacting electron pose a lot of interesting questions. For example:

1. The general dispersion law (72) may be related to the problem
of the lepton generations (electron, muon, tauon). Detailed numerical analysis
should give reply on this question.

2. The generation of an electromagnetic-like field by the coset transformations of
manifold $CP(3)$ is also an interesting question.

3. The self-interacting electron is sharply concentrated in the the area with linear
size of the order of Compton wave-length. A new calculation of the Lamb shift
in the framework of non-local electron should be done. It
may avoid divergences without renormalization procedure.

These acute old problems could not be solved separately and they require some general framework. Here I would like to give an outlook (without technical details given above in ``Conclusion") of my approach to these topics.

1. My main goal is to build a simple unification of quantum and relativity principles. This means that a minimal number of primordial elements and postulates should be used and only standard mathematical operations are eligible. In my case only rays of quantum states are primordial elements and the assumption about the unitary character of their dynamics have been used. But these simplest assumptions lead to deep reconstruction in the spirit of ``Deterministic underlying theory: We suspect that our world can be understood by starting from a pre-quantized classical, or 'ontological', system" \cite{'t Hooft2}. In my model the deterministic underlying theory is rooted in the complex projective Hilbert space $CP(N-1)$.

2. The most fundamental difficulty, in fact the stumbling-block on the way of unification of quantum and relativity principles, is the \emph{localization problem} \cite{Horwitz,Jones1,Jones2}. This problem is so acute that it evokes a new concept of space-time as some sub-manifold of a Hilbert space \cite{Kryukov1} and a new principle of ``functional relativity" in this space \cite{Kryukov2}.

Localization being treated as the ability of a coordinate description of an object in classical relativity is closely connected with operational identification of ``events" \cite{Einstein1}. It is tacitly assumed that all classical objects (frequently represented by material points) are self-identical and they can not disappear because of the energy-momentum conservation law. However the quantum identification and therefore localization of particles cannot be done in a similar manner (like in special relativity) and it requires a physically motivated operational procedure with corresponding mathematical description. In order to do it some conservation law in the state space expressing the ``self-identification" or ``self-conservation" should be formulated \cite{Le1,Le2,Le3}.

It turns out that QFT dictates the necessity to reformulate QM according to a new principle of invariance as well as electromagnetic theory insisted to reformulate  Newton's kinematic and dynamics in a relativistic manner. This is the true reason why I called this theory ``super-relativity". \emph{Formally it means that the complex projective geometry lies in the base of quantum theory so that distance between quantum states replaces a distance between events in space-time. }

3. Two simple observations I've put in the basis of my theory:

A. Quantum interference phenomenon shows the symmetry of relative space-time transformations of a whole setup. These have been studied in ordinary quantum theory. Such symmetries reflect, say, the \emph{first order of relativity}:
the physics is the same for any \emph{complete setup} subject (kinematical, not
dynamical!) to shifts, rotations, boosts as a whole in a single Minkowski space-time.

B. There is however a different type of tacitly assumed symmetry that may be formulated on the intuitive level as invariance of the physical properties of ``quantum particles" , i.e. the invariance of their quantum numbers like mass, spin, charge, etc., relative variation of the quantum amplitude. Say, physical properties of electrons are the same in two setups $S_1$ and $S_2$. I postulated that the invariant content of these properties may be kept if one makes the infinitesimal variation of some ``flexible quantum setup"  reached by small variation of some fields or adjustment of tuning devices; it comprises the \emph{the second order of relativity or ``super-relativity"} \cite{Le1,Le2,Le3}.

This ``flexible quantum setup" is an invariant construction (in the sense of Fubini-Study metric) since it is built from the tangent vector fields on a $CP(N-1)$ manifold represented by the generators of $SU(N)$ expressed in the local coordinates $\pi^i$. These generators are state-dependent dynamical variables \cite{Le4} similar to those which arose in Weinberg's attempted generalization of the ordinary QM \cite{Weinberg}. The non-bilinear character of corresponding observables is naturally provided here by the state-dependent character of the local dynamical variables (LDV's).

It is possible to say that ``super-relativity" is a local version of the ``functional relativity" under some reservations in the choice of Hilbert space and the space-time construction.

4. There is a different approach to extended quantum objects studying in the framework of Thermo Field Dynamics (TFD). These arise due to boson condensation providing the dynamical reconstruction of symmetry \cite{TFD}. Dynamical reconstruction of $SU(N)$ symmetry has been already discussed in the context of projective representations of $SU(N)$ generators \cite{Le5}.
TFD assumes the applicability of the second quantization scheme. It has been shown by Blochintzev \cite{Bl1,Bl2} that there is an essential problem in the application of this method for strongly interacting and self-interacting fields.

Since I investigate the self-interacting electron it would be reasonable to start from its relativistic quantum (not secondly quantized) model (according to ``A") and to study their quantum invariants under infinitesimal deformations of amplitudes (according to ``B"). I used for this purpose Dirac's electron and the geometric features of the Foldy-Wouthuysen unitary transformations from $SU(4)$.

5.  The main goal was to get new non-linear wave equations and to study its lump (soliton-like) solution for the ``field shell" associated with the surrounding field of single electron. Dirac's equations are the first order linear system of PDE's. The closest to such equations are quasi-linear first order PDE's, the theory of which is properly developed \cite{Courant}. These equations very naturally follow from the conservation law of the energy momentum field expressed by the affine parallel transport of the energy momentum vector field in $CP(3)$, and agrees with the Fubini-Study metric \cite{Le1}. Since all quantum dynamics concentrated now in $CP(3)$ base manifold and dynamical space-time arises merely for the parametrization of section in the tangent fibre bundle, the differentiation in complex local coordinates $\pi^i$ has been used instead of a variation procedure.

\begin{acknowledgements}
I am sincerely grateful to Larry Horwitz for interesting discussions and essential
improvements of English.
\end{acknowledgements}

\end{document}